\documentclass[letterpaper,journal]{IEEEtran}
\usepackage{amsmath,amsfonts}
\usepackage{algorithmic}
\usepackage{algorithm}
\floatstyle{ruled}
\restylefloat{algorithm}
\usepackage{array}

\usepackage[caption=false,font=normalsize,labelfont=sf,textfont=sf]{subfig}
\usepackage{textcomp}
\usepackage{url}
\usepackage{verbatim}
\usepackage{graphicx}
\usepackage{cite}
\usepackage{siunitx} 
\usepackage{multirow}
\usepackage{bbm}
\usepackage{flushend}
\usepackage{dblfloatfix}

\makeatletter
\renewcommand{\subsubsection}{\@startsection{subsubsection}{3}{\z@}%
  {1.2ex plus 0.3ex minus 0.2ex}%
  {0.45ex plus 0.1ex}%
  {\normalfont\normalsize\itshape}}
\makeatother
\begin{document}

\title{Update for Decisions, Not Freshness: Goal-Oriented Status Updating and Selective Offloading at the Network Edge}

\author{Jianpeng Qi, Qiyang Zhang,
Chao Liu, \IEEEmembership{Member, IEEE}, Jing Sun, Yimei Liu,
Yanwei Yu, \IEEEmembership{Member, IEEE},
Yingjie Wang, \IEEEmembership{Member, IEEE}, and
Wei Ni, \IEEEmembership{Fellow, IEEE}
}

\markboth{IEEE Transactions on Mobile Computing,~Vol.~XX, No.~X, Month~YYYY}%
{Qi \MakeLowercase{\textit{et al.}}: Goal-Oriented Status Updating and Selective Offloading at the Network Edge}


\maketitle

\begin{abstract}
In an edge--cloud collaborative edge-computing environment, an edge node (EN)
must decide whether each user task should be executed locally, forwarded to a
remote service (or cloud) node (SN), or rejected. The EN observes its local state directly but receives the SN state
only through an intermittently refreshed cache. Status updating and task
control therefore form an asynchronous closed loop under partial observability. Freshness-driven schemes, including those based on Age of
Information (AoI), do not directly value an update by its effect on subsequent
task decisions. We propose \textsc{CoSMO} (Co-design of Semantic-state
Management and Offloading), a cooperative event-driven reinforcement learning
(RL) framework that coordinates semantic status management and selective
offloading through realized task utility.
\textsc{CoSMO} learns a compact representation of the heterogeneous SN service
state. At the SN, a recurrent semi-Markov double deep Q-network (Double DQN)
agent jointly selects
send/no-send and the next decision interval. At the EN, a task-terminal
off-policy value-learning agent makes hierarchical gate--route decisions from
local observations and stale remote semantics. The agents maintain separate
observations and value targets but share the same realized task-utility stream,
without centralized execution. Across the evaluated workload families,
\textsc{CoSMO}'s reported relative improvement in on-time completion rate over
the best-performing competing method averages $18.6\%$--$21.2\%$. For
capacity-aware decision accuracy across the three strict-overload points, the
corresponding reported gains average $17.6\%$--$17.9\%$.
\end{abstract}

\begin{IEEEkeywords}
Mobile edge computing, computation offloading,
goal-oriented status updating, age of information, deep reinforcement
learning.
\end{IEEEkeywords}

\section{Introduction}
\label{sec:intro}

\IEEEPARstart{E}{dge} computing increasingly relies on edge--cloud collaborative
execution~\cite{mec_survey}. Each arriving task must be assigned to a local or
remote service node that can process it within an acceptable delay. This
assignment requires the dispatcher to compare local and remote service status,
including service availability, queue backlog, and expected service delay.
Local and remote service states differ in observability. The dispatcher
observes its local state directly, and obtains remote state through a
capacity-limited information link. Because this link cannot support continuous
synchronization, the dispatcher caches the latest remote status. The cached
view becomes stale as the remote service state evolves. The resulting
stale-status dispatch problem requires task decisions based on incomplete and
potentially outdated remote information.

\begin{figure}[t]
\centering
\includegraphics[width=\columnwidth]{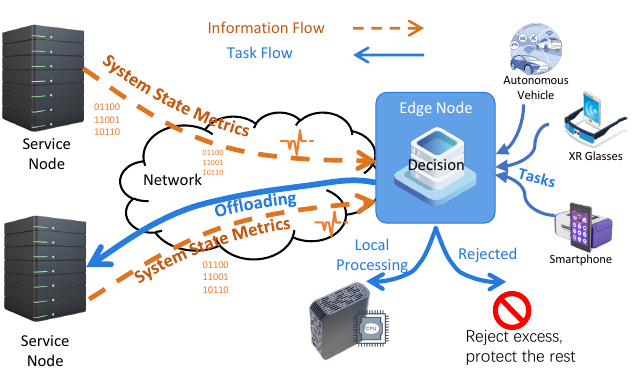}
\caption{Information and task flows in goal-oriented status updating for CFN.}
\label{fig:scene}
\end{figure}

This local--remote dispatch setting has been studied extensively in mobile
edge computing (MEC)~\cite{11278072}. Compute awareness in Compute-First
Networking (CFN) and Computing Power Network (CPN) architectures provides a
concrete instance of the stale-status problem~\cite{cfn_ton,cfn_anycast,mucvr}. As depicted in \figurename~\ref{fig:scene}, user equipments (UEs) or end devices submit
tasks to an edge node (EN), which acts as the dispatcher. The EN observes its
own compute and queue state directly. By contrast, a remote service node (SN)
is represented only by cached service status delivered over a separate
information flow. For each arriving task, the EN chooses local execution,
offloading to the SN, or rejection. Under overload (offered-load ratio
$\rho>1$), accepting every task can compromise the service quality of tasks that
the system can serve. The EN therefore makes each choice from local
observations and a potentially outdated SN view obtained through intermittent
status updates~\cite{cpn_learning}. The joint problem is to determine
\emph{when} the SN should refresh its status and \emph{how} the EN should act
on the cached information.

Existing status-update methods commonly characterize update value through
freshness. The age of information (AoI) and its variants quantify how outdated
the receiver's knowledge is and support status-update scheduling and
monitoring~\cite{aoi_survey,query_aoi,age_sem_importance}. More recently,
goal-oriented and semantics-aware communication has shifted the focus from
faithful source reproduction to information relevant to the receiver's
task~\cite{data_significance,goal_tensor,goal_metrics,semaware_source,
freshness_to_semantics}. The age of incorrect information provides a
complementary criterion by measuring how long the cached value is wrong~\cite{aoii,aoii_delay}. These approaches provide distinct criteria for
determining which status information to transmit.

In the compute-aware service-offloading setting considered here, the update value
depends on the downstream task decision rather than on freshness or state
deviation alone~\cite{11396674}. A fresh status, or one that differs markedly
from its cached copy, may still produce the same local/offload/reject action.
Conversely, a small service-state change may alter the selected action. This
gap motivates goal-oriented status updating: The SN should refresh its status
when the new information is likely to improve downstream task handling, rather
than only when the cached status is outdated or different.

Realizing this goal-oriented view presents three challenges. First, the EN and
SN operate under \textit{partial observability and information asymmetry}. The
EN makes each task decision from a cached SN status and never observes the SN's
instantaneous backlog. Meanwhile, the SN schedules updates without observing
the EN's local arrivals or queue. Neither of the agents has the complete
information required to evaluate the joint decision.

Second, updating and task decisions form \textit{a bidirectionally coupled
closed loop}, although they are triggered by different events. Each update influences
EN task decisions until the next refresh. These decisions reshape the SN load
and, in turn, affect when another update is useful. Optimizing either decision
in isolation may therefore omit this feedback.

Third, feedback arrives at \textit{two event horizons}. The utility of an EN
action becomes known only when the corresponding task reaches a terminal
outcome. An SN update is instead evaluated from task utilities accumulated over
the following decision interval and from the value of later intervals. The
learning design must align a task-terminal EN return with a semi-Markov SN
return without imposing a common artificial clock.

In this paper, we formulate this asynchronous coupling as the control problem and propose
\textsc{CoSMO}, a cooperative semantic-aware reinforcement learning (RL)
framework. At each SN decision
epoch, the SN maps its local service state into a compact representation and
 selects whether to transmit it and when to decide again. At each task
arrival, the EN selects local execution, offloading, or rejection from its
local observation and the latest cached semantic. The agents retain separate
observations and value targets, while their rewards are derived from a common stream
of realized task utilities. This event-driven coordination links what is
transmitted, when it is transmitted, and how the EN uses the cached information
for accept/reject and routing decisions.

The contributions of this paper are summarized as follows.
\begin{itemize}
\item We formulate stale-state edge offloading as an asynchronous closed-loop
  control problem with two event clocks: SN status decisions and EN task
  arrivals. This formulation includes active rejection under overload and
  evaluates information by downstream task utility rather than freshness alone.
\item For the SN status-update decision, we design a temporal semantic decision
  network that uses cached semantic-state information, update age, and recent
  task feedback. A gated recurrent unit (GRU)-based dueling Double deep
  Q-network (Double DQN) head selects whether to transmit a semantic
  update and the next decision interval. A semi-Markov target uses
  interval-level rewards aggregated from downstream task utilities and update
  cost.
\item For the EN task-forwarding decision, we design a dual-branch state encoder
  and a hierarchical gate--route decision network. The gate first determines
  whether an arriving task should be accepted or rejected. The route head
  then selects local execution or SN offloading for accepted tasks. This
  hierarchy makes active rejection explicit under overload. The EN remains an
  off-policy RL agent, but its delayed task outcome forms a
  task-terminal value target rather than an inter-task bootstrap.
\item We evaluate the complete \textsc{CoSMO} framework across a range of load
  conditions and two arrival processes in ns-3. With the controlled
  single-EN/single-SN topology, we show end-to-end feasibility
  and improvements in on-time completion and capacity-aware decision accuracy,
  together with the status-update cost tradeoff relative to the compared
  freshness-driven and fixed-rule policies.
\end{itemize}

The remainder of this paper is organized as follows.
Section~\ref{sec:model} presents the system model and problem formulation.
Section~\ref{sec:algorithm} elaborates on the goal-oriented co-design and its
training procedure. Section~\ref{sec:experiments} describes the experiments
and results. The paper concludes with the limitations and
directions for future work in Section \ref{sec:conclusion}.

\section{Related Work}
\label{sec:related}
We organize related work into three areas: Compute-aware service offloading,
freshness- and goal-oriented status updating, and semantic-aware edge
offloading. The first two areas separately address task assignment and remote
state refresh. The third connects these decisions and motivates their joint
design under stale information and overload.

\subsection{Compute-Aware Service Offloading}
CFN jointly schedules computation and connectivity to serve tasks across cloud,
data-center, and edge tiers~\cite{cfn_ton,mec_survey}. Existing studies examine
anycast-based service offloading in software-defined computing power
networks~\cite{cfn_anycast}, online scheduling of inference services with
learning~\cite{cpn_learning}, and edge-enabled multi-user collaboration for
interactive applications~\cite{mucvr}. Load-aware multi-agent offloading partitions
tasks among local devices, mobile LEO satellites, and a remote cloud data
center based on current satellite resource and load
states~\cite{XIAO2026112082}.  These studies largely assume that the
controller or EN has sufficiently accurate node status and therefore focus on
offloading or placement. In contrast, this paper considers per-task decisions based on stale,
cached SN status. This setting requires the SN refresh schedule and EN task
policy to be coordinated.

\subsection{Freshness and Goal-Oriented Updating}

A substantial body of work quantifies update utility through information
freshness. AoI and its variants support update scheduling and
monitoring~\cite{aoi_survey}, including pull-based or query-driven
settings~\cite{query_aoi} and multi-source scheduling weighted by semantic
importance~\cite{age_sem_importance}. Because freshness does not indicate
whether a cached state is correct, the age of incorrect information measures
how long that state remains wrong~\cite{aoii,aoii_delay}. Goal-oriented and
semantics-aware communication moves closer to the receiver's objective by
prioritizing information relevant to the downstream
task~\cite{data_significance,goal_tensor,goal_metrics,semaware_source,
goal_tracking,freshness_to_semantics}. Our earlier AVA work combined value of
information (VoI) and AoI to decide whether to update raw service status for
subsequent task forwarding~\cite{11396674}.

The present work changes this control boundary in three coupled respects. It
learns a compact dynamics-aware status abstraction instead of transmitting the
full raw state. It gives the SN a paired send/no-send and
next-decision-interval action. It also extends EN control to three-way task
handling under overload through local execution, offloading, or active
rejection. To this end, CoSMO coordinates status abstraction, update timing,
overload-aware task acceptance, and execution routing through realized
downstream task utility.

\subsection{Semantic-Aware Edge Offloading}

Semantic and task-oriented methods are increasingly applied to edge
computing and offloading. Surveys summarize semantic edge computing and
its role in prospective 6G networks~\cite{sem_edge_survey}. On the control
side, deep reinforcement learning (DRL) is widely used for computation
offloading and resource allocation. Lyapunov-guided DRL stabilizes online
offloading under stochastic arrivals~\cite{drl_lyapunov_offload}. Other
DRL-based schemes allocate resources for task-oriented semantic
communication~\cite{drl_semantic_resource} and semantic-aware networks with
task offloading~\cite{sem_task_offload}. Multi-objective RL jointly optimizes semantic extraction and task scheduling
in vehicular edge computing~\cite{lin2026sjtsra}. QoE-driven MAPPO jointly optimizes semantic extraction and multi-task
offloading for multiple users connected to a single access point and a
single edge server~\cite{11386889}. Utility-loss minimization schedules semantic status generation and satellite
transmission to balance state-estimation mismatch and energy
consumption~\cite{huang2025utility}. These approaches typically optimize
offloading or resource allocation for an available state and emphasize
throughput or latency. In contrast, we couple offloading with the upstream
decision of when to update its remote state. The policy can also decline tasks
when the available capacity cannot support timely completion.

In summary, prior work provides the individual foundations for compute-aware
offloading, goal-oriented updating, semantic representation, and learning-based
edge control. CoSMO differs by composing them into one asynchronous decision
process. Particularly, a recurrent semi-Markov updater controls both transmission and revisit
time, while a task-terminal value learner uses the resulting cached semantic
for overload-aware accept/reject decisions and route selection. This scope
differs from freshness optimization with a fixed downstream controller and from
offloading optimization that treats the available remote state as given.

\section{Decision-Oriented Closed-Loop System Model and Problem Formulation}
\label{sec:model}

\subsection{System Model}
In this section, we specify the three-tier MEC architecture, task and mobility
models, node service states, agent observations, event-driven decisions, and
optimization objective.

\subsubsection{Network Architecture}
We consider a three-tier MEC system consisting of multiple UEs, an EN, and
an SN. The UEs submit tasks to the EN over wireless uplinks. The EN
acts as the task controller and, for each arriving task, either processes it
locally, offloads it to the SN for execution, or rejects it. The EN and SN are
connected by a backhaul link used for task upload, result return, and SN status
updates. As shown in \figurename~\ref{fig:scene}, the system contains two flows:
An \emph{information flow}, in which the SN sends its state information to the EN,
and a \emph{task flow}, which carries task arrivals, offloaded tasks, and
returned results.

The SN status updates are delivered intermittently to the EN and stored with their
timestamps in an EN-side cache. At a task arrival, the EN observes local EN-side
information and uses the latest cached SN status, which may differ from the
instantaneous SN service state. After receiving and caching a status update, the
EN can notify the SN through an acknowledgment (ACK) mechanism. A rejected task
terminates at the EN and is not forwarded to either of the execution queues. The UE
mobility changes the wireless access condition and connection persistence,
which are represented by the access-state variables defined below.

\subsubsection{Task Model and Delay Cost}

We assume that tasks from the UEs arrive at the EN according to a stochastic process. The
$k$-th task is characterized by the following five-tuple
\begin{equation}
J_k = (t_k, i_k, b_k, c_k, D_k),
\end{equation}
where $t_k \in \mathbb{R}_{\ge 0}$ is the arrival time, and
$i_k \in \{1,2,\ldots,N\}$ identifies the source UE among the $N$ UEs. The
variables $b_k,c_k,D_k \in \mathbb{R}_{>0}$ denote the input data size, compute
demand, and delay constraint, respectively.

For delay accounting, the EN-side route indicator $a_k$ applies only to the accepted
tasks: $a_k=0$ denotes local execution at the EN, and $a_k=1$ denotes offloading
to the SN. The complete EN action space and the SN update action are specified
below. A locally executed task enters the EN queue, whereas an offloaded task is
executed at the SN and incurs EN--SN upload and return delays. At either node, the 
accepted tasks enter service, based on the first-come first-served (FCFS)
ordering. For an accepted task ($a_k \in \{0,1\}$), the end-to-end completion
delay is
\begin{equation}
T_k = T_k^{\mathrm{ul}} + a_k \bigl(T_k^{\mathrm{up}} + T_k^{\mathrm{dn}}\bigr)
+ T_k^{\mathrm{wait}} + T_k^{\mathrm{comp}},
\end{equation}
where $T_k^{\mathrm{ul}}$ is the wireless uplink delay from UE $i_k$ to the EN,
$T_k^{\mathrm{up}}$ is the EN-to-SN task upload delay, $T_k^{\mathrm{dn}}$ is the
SN-to-EN result download delay, $T_k^{\mathrm{wait}}$ is the queueing delay at
the selected node, and $T_k^{\mathrm{comp}}$ is its computation delay.

We do not embed an explicit queueing model in the optimization problem. Instead,
$T_k^{\mathrm{wait}}$ and $T_k^{\mathrm{comp}}$ are realized delays generated by
the execution process and enter $T_k$ directly. An accepted task succeeds if
and only if $T_k \le D_k$. The node service states represent the
corresponding queue and workload conditions. An actively rejected task
terminates immediately without consuming compute or transmission resources and
is scored through the rejection outcomes in the task-utility definition.

\subsubsection{Mobility and Access State}

In the considered single-EN setting, mobility affects task performance primarily through the
wireless access process. UE motion relative to the EN changes link quality,
task-upload delay, and link sustainability, thus affecting the EN's
offloading choice~\cite{mec_survey,mec_arch_offload}. Following mobility-aware
vehicular edge computing (VEC) offloading models~\cite{sun2024bargain}, we use a
compact access-state representation rather than the complete UE trajectory.

For task $J_k$ from UE $i_k$ arriving at time $t_k$, the corresponding mobility
state is
\begin{equation}
\mathbf{m}_k
=
\bigl(
R_{i_k}^{\mathrm{ul}}(t_k),\,
s_{i_k,k},\,
\tau_{i_k,k}
\bigr),
\end{equation}
where $R_{i_k}^{\mathrm{ul}}(t_k)$ is the estimated effective uplink throughput
observed at arrival, $s_{i_k,k}$ is the access-state trend, and
$\tau_{i_k,k}$ is the residual connection time. These three quantities
characterize the current access capability, short-term direction of change, and
connection sustainability, respectively.

The uplink-throughput feature is derived from measured task-upload delay. For a
task with data size $b_k$ and measured UE--EN uplink delay
$T_k^{\mathrm{ul}}$, the measured uplink-throughput sample is
\begin{equation}
\hat R_{i_k}^{\mathrm{ul}}(t_k)
=
\frac{b_k}{T_k^{\mathrm{ul}}}.
\end{equation}
To reduce measurement noise, the observation uses an exponentially weighted
moving average (EWMA). This recursive estimator is also used for network-delay
estimation, including TCP's smoothed-RTT estimator~\cite{rfc6298}:
\begin{equation}
R_{i_k}^{\mathrm{ul}}(t_k)
=
\alpha_R \hat R_{i_k}^{\mathrm{ul}}(t_k)
+
(1-\alpha_R)R_{i_k}^{\mathrm{ul}}(t_{k^-}),
\end{equation}
where $\alpha_R \in [0,1]$ is the smoothing coefficient, and $t_{k^-}$ denotes
the previous task-arrival time of the same UE. The trend feature is then given as
\begin{equation}
s_{i_k,k}
=
R_{i_k}^{\mathrm{ul}}(t_k)
-
R_{i_k}^{\mathrm{ul}}(t_{k^-}).
\end{equation}
The residual connection time $\tau_{i_k,k}$ is computed from the relative
positions and velocitys of UE $i_k$ and the EN. It represents the remaining time
before the UE leaves the configured service radius.

\subsubsection{Node Service State}
To support local decisions, each execution node
$r\in\{\mathrm{EN},\mathrm{SN}\}$ maintains a service state:
\begin{equation}
\mathbf{x}_t^r
=
\bigl(
N_t^r,\,
W_t^r,\,
B_t^r,\,
M_t^r
\bigr),
\qquad r\in\{\mathrm{EN},\mathrm{SN}\}.
\end{equation}
Here, $N_t^r$ is the idle compute resource, $W_t^r$ is the head-of-line waiting
time in the waiting queue, and $B_t^r$ is the total residual compute demand of
outstanding tasks. The variable $M_t^r$ denotes the available storage. The
workload component is then given by
\begin{equation}
B_t^r
=
\sum_{j \in \mathcal{Q}_t^{\mathrm{wait},r}} c_j
+
\sum_{j \in \mathcal{Q}_t^{\mathrm{run},r}} c_j^{\mathrm{rem}}(t),
\qquad r\in\{\mathrm{EN},\mathrm{SN}\},
\end{equation}
where $\mathcal{Q}_t^{\mathrm{wait},r}$ and $\mathcal{Q}_t^{\mathrm{run},r}$ are
the waiting and running task sets at node $r$, respectively; $c_j$ is the total compute demand
of task $j$; and $c_j^{\mathrm{rem}}(t)$ is its residual demand at time $t$.

\subsection{Decision-Oriented Status and Observations}

We next define the SN status abstraction transmitted to the EN, the observations
available to each agent, and the cached remote status used by the EN.

\subsubsection{SN Status Abstraction}
The EN relies on delayed SN information, while frequent raw-state transmission
incurs communication overhead and may retain details unnecessary for task
decisions. We encode the information-flow payload as a compact SN
status semantic
\begin{equation}\label{eq:semantic-encoder}
\mathbf{z}_t = f_{\theta}\bigl(\mathbf{x}_t^{\mathrm{SN}}\bigr),
\qquad
f_{\theta}:\mathbb{R}^{4}\!\to\!\mathbb{R}^{2},
\end{equation}
where $f_{\theta}(\cdot)$ is a parameterized status-abstraction mapping, and
$\mathbf{z}_t \in \mathbb{R}^{2}$ is the resulting status semantic. The SN
transmits this semantic to the EN, which caches it for subsequent task
decisions. Section~\ref{sec:algorithm} specifies $f_{\theta}(\cdot)$.

\subsubsection{Asymmetric EN/SN-Side Observations}

The system operates under partial observability and information asymmetry.
During online execution, the SN and EN only use their locally available
observations.

\emph{SN observation at decision epoch $u_n$:} Let $u_n$ denote the $n$-th SN decision epoch, and
$\sigma_n\in\{0,1\}$ indicate whether the SN transmits a semantic update at
that epoch. The SN information set at $u_n$ is defined as
\begin{equation}\label{eq:sn_info_set}
\mathcal{I}_n^{\mathrm{SN}} = \bigl\{
    \mathbf{z}_{u_n},\,
    \mathbf{z}^c_{\nu_n},\,
    A_{u_n},\,
    \mathcal{H}_n^{\mathrm{comp}},\,
    \ell_{u_n}^{\mathrm{SN}}
\bigr\},
\end{equation}
where $\mathbf{z}_{u_n}=f_{\theta}(\mathbf{x}_{u_n}^{\mathrm{SN}})\in\mathbb{R}^{2}$
is the status semantic of the current service state
$\mathbf{x}_{u_n}^{\mathrm{SN}}\in\mathbb{R}^{4}$.
Let $\nu_n\leq u_n$ be the source timestamp of the status semantic most
recently acknowledged by the EN before decision epoch $u_n$. Then
$\mathbf{z}^c_{\nu_n}$ is that cached copy and
$A_{u_n}=u_n-\nu_n$ is its AoI. This definition remains valid when one or more
preceding SN decisions choose no-send. Here
$\mathcal{H}_n^{\mathrm{comp}}$ is the history of offloaded tasks completed by
$u_n$. The variable $\ell_{u_n}^{\mathrm{SN}}$ is a local statistic of the
SN-to-EN update link, such as its mean rate, rate standard deviation, and
short-term rate trend.

Task outcomes, including success and final delay, become observable to the SN
only after completion. Hence, $\mathcal{H}_n^{\mathrm{comp}}$ contains only
tasks offloaded before $u_n$ whose outcomes are known:
\begin{equation}
\mathcal{H}_n^{\mathrm{comp}}
\subseteq
\left\{
    \begin{aligned}
        &(J_i, a_i, T_i):\ t_i < u_n,\ a_i = 1,
        & t_i + T_i \le u_n
    \end{aligned}
\right\}.
\end{equation}

\emph{EN observation at task-arrival epoch $t_k$:} The EN information set
at the arrival time of the $k$-th task is defined as
\begin{equation}\label{eq:ap_info_set}
\mathcal{I}_k^{\mathrm{EN}} =
\bigl\{
    \mathbf{m}_k,
    \mathbf{x}_{t_k}^{\mathrm{EN}},
    J_k,
    \mathbf{z}^c_{u_{n(k)}},
    A_{t_k},
    \ell_{t_k}^{\mathrm{EN}}
\bigr\},
\end{equation}
where $\mathbf{m}_k$ is the source UE's mobility state,
$\mathbf{x}_{t_k}^{\mathrm{EN}}$ is the local EN service state, and $J_k$
contains the current task features. The variable
$\mathbf{z}^c_{u_{n(k)}}$ is the SN status semantic cached at the EN, and
$A_{t_k}=t_k-u_{n(k)}$ is its age. Here, $u_{n(k)}$ is the source timestamp of
the latest SN update cached before $t_k$. Finally,
$\ell_{t_k}^{\mathrm{EN}}$ is a local statistic of the EN-to-SN link, such as
its mean rate, rate standard deviation, and short-term rate trend.


\subsection{Closed-Loop Event-Driven Decision Process}

The closed-loop decision process evolves in continuous time through two event
streams. Task arrivals $\{t_k\}$ trigger EN task decisions, whereas SN decision
epochs $\{u_n\}$ trigger semantic-update decisions. Because task interarrival
times and SN decision intervals are generally nonuniform, we model the control
problem as an asynchronous event-driven semi-Markov decision process
(semi-MDP). The two event-triggered decisions are defined below.

\emph{Task-offloading decision:} At each arrival time $t_k$, the EN observes
the local information set $\mathcal{I}_k^{\mathrm{EN}}$ and chooses a task
action from the three-way action space
\begin{equation}\label{eq:en_action_space}
\mathcal{A}^{\mathrm{EN}} = \{0,1,2\},
\qquad
a_k \in \mathcal{A}^{\mathrm{EN}},
\end{equation}
where the action indices are defined by
\begin{equation}\label{eq:en_action_semantics}
a_k = \begin{cases}
    0, & \text{local execution at the EN},\\
    1, & \text{offloading to the SN},\\
    2, & \text{active rejection}.
\end{cases}
\end{equation}
Equivalently, the EN policy is written as
\begin{equation}
a_k = \pi^{\mathrm{EN}}\!\bigl(\mathcal{I}_k^{\mathrm{EN}}\bigr),
\end{equation}
where $a_k\in\mathcal{A}^{\mathrm{EN}}$ follows the action semantics
in~\eqref{eq:en_action_semantics}.

\emph{State-update decision:} At the $n$th SN decision epoch $u_n$, the SN
controls the information flow by jointly deciding whether to transmit the
current status semantic and when to make the next decision:
\begin{equation}
(\sigma_n,g_n) = \pi^{\mathrm{SN}}\!\bigl(\mathcal{I}_n^{\mathrm{SN}}\bigr),
\end{equation}
where $\sigma_n\in\{0,1\}$ is the transmit indicator for
$\mathbf{z}_{u_n}$, and $g_n\in\mathcal{G}\subseteq[g_{\min},g_{\max}]$ is the
waiting time before the next decision epoch. The next epoch is therefore written as
\begin{equation}
u_{n+1} = u_n + g_n.
\end{equation}
If $\sigma_n=1$, the EN cache is updated to $\mathbf{z}_{u_n}$ with timestamp
$u_n$ once the EN receives it. If $\sigma_n=0$, the SN skips the current
transmission opportunity and revisits the update decision after $g_n$ time units
at $u_{n+1}$; during this interval, the EN keeps the previous cached semantic
and timestamp.

The two decisions therefore operate on distinct event streams. The EN acts at
each task arrival, whereas the SN selects semantic transmission and the next
interval at each SN decision epoch. Define the number of EN task decisions in the
$n$-th SN interval $[u_n,u_{n+1})$ as
\begin{equation}
N_n = \left|
    \left\{
      k: u_n \le t_k < u_{n+1}
    \right\}
\right|,
\end{equation}
whose mean satisfies
\begin{equation}
\bar N = \mathbb{E}[N_n]
\approx \frac{\bar g}{\bar\Delta},
\end{equation}
where $\bar g=\mathbb{E}[g_n]$ is the mean SN decision interval and
$\bar\Delta=\mathbb{E}[t_{k+1}-t_k]$ the mean task inter-arrival time.
The formulation does not require $\bar g \gg \bar\Delta$. When
$\bar g \gg \bar\Delta$, the system behaves as a two-timescale process, and one
update typically affects many tasks. When $\bar g$ is comparable to or smaller
than $\bar\Delta$, $N_n$ may be small, but the model and training pipeline remain
unchanged. A transmitted update can therefore affect EN task decisions
throughout the subsequent interval.

\subsection{Problem Formulation}

The objective is to jointly design the SN semantic-update policy
$\pi^{\mathrm{SN}}$ and the EN task-offloading policy $\pi^{\mathrm{EN}}$ to
maximize the long-term average net system utility while balancing task utility
against semantic-update cost.
Suppose that the $k$-th task arrive at $t_k$, and let the $n$-th SN decision epoch occur at
$u_n$. The objective is written as
\begin{equation}\label{eq:overall_obj}
\max_{\pi^{\mathrm{SN}},\,\pi^{\mathrm{EN}}}
\lim_{T \to \infty} \frac{1}{T} \mathbb{E} \left[
    \sum_{k:t_k \le T} U_k - \lambda_s \sum_{n:u_n \le T} C_n
\right],
\end{equation}
where $U_k$ is the task utility defined below, $C_n$ is the update
communication cost at SN decision epoch $u_n$, and $\lambda_s>0$ is the
update-cost coefficient. To specify $C_n$, let $\sigma_n\in\{0,1\}$ indicate
whether the SN transmits the current status semantic at $u_n$, let $c_s$ be the
per-dimension communication cost, and $d_z$ be the semantic dimension. Then,
\begin{equation}
C_n = \sigma_n c_s d_z.
\end{equation}
Thus, a no-send decision ($\sigma_n=0$) incurs zero update cost, while a send
decision ($\sigma_n=1$) is charged only for transmitting the current status
semantic.

For each task, the decision-oriented utility $U_k$ measures the contribution of
the EN decision after its outcome is known. It combines an outcome-dependent
value with the task-side execution cost, i.e.,
\begin{equation}
\Phi_k = \lambda_T T_k + \lambda_c \mathbbm{1}_{\{a_k=1\}} b_k ,
\end{equation}
where $T_k$ is the realized end-to-end delay; $b_k$ is the task input size; and
$\lambda_T$ and $\lambda_c\ge 0$ are the delay and EN--SN transmission coefficients,
respectively. The scaling weight $\eta_c\ge0$ controls how strongly this
execution cost enters the decision-oriented reward.
Let $\chi_k$ denote the resulting closed-loop decision outcome:
\begin{itemize}
\item \emph{True positive} ($\mathrm{TP}$): The EN chooses local execution or
  offloading, the task is accepted, and it completes by its deadline.
\item \emph{False positive} ($\mathrm{FP}$): The EN chooses local execution or
  offloading, but the selected node cannot accept the task or the accepted task
  misses its deadline.
\item \emph{True negative} ($\mathrm{TN}$): The EN actively rejects the task and
  no deadline-feasible execution route exists under the current estimated system
  state.
\item \emph{False negative} ($\mathrm{FN}$): The EN actively rejects the task
  even though at least one deadline-feasible execution route exists.
\end{itemize}
\begingroup
\setlength{\arraycolsep}{2pt}
\begin{equation}\label{eq:task_utility}
U_k = \begin{cases}
    w_{\mathrm{TP}} V - \eta_c \Phi_k, & \chi_k=\mathrm{TP};\\[2pt]
    -w_{\mathrm{FP}} V - \eta_c \Phi_k, & \chi_k=\mathrm{FP};\\[2pt]
    w_{\mathrm{TN}} V, & \chi_k=\mathrm{TN};\\[2pt]
    -w_{\mathrm{FN}} V, & \chi_k=\mathrm{FN};\\[2pt]
    0, & \text{no EN decision for task } k.
\end{cases}
\end{equation}
\endgroup
The scalar $V>0$ sets the base task value, and the positive coefficients
$w_\chi$, one for each $\chi\in\{\mathrm{TP},\mathrm{FP},\mathrm{TN},
\mathrm{FN}\}$, weight the corresponding decision outcomes. Thus, $U_k$ rewards
successful acceptance and correct rejection, and penalizes
accepted failures and erroneous rejection.

The optimization is subject to the following constraints.

\textbf{C1. Resource constraints:} For each node
$r\in\{\mathrm{EN},\mathrm{SN}\}$ and all time $t$,
\begin{equation}
\begin{aligned}
&0 \le N_t^{r} \le N_r^{\max};\\
&0 \le M_t^{r} \le M_r^{\max};\\
&|\mathcal{Q}_t^{r}| \le Q_r^{\max},
\end{aligned}
\end{equation}
with the SN provisioned more strongly than the EN:
\begin{equation}
\begin{aligned}
N_{\mathrm{SN}}^{\max} &> N_{\mathrm{EN}}^{\max};\\
M_{\mathrm{SN}}^{\max} &> M_{\mathrm{EN}}^{\max}. \\
\end{aligned}
\end{equation}

\textbf{C2. Task-outcome mechanism constraint:}
Let $\rho(0)=\mathrm{EN}$ and $\rho(1)=\mathrm{SN}$. Before inserting task $k$,
the feasibility of accepting it on route $a\in\{0,1\}$ is
\begin{equation}
\mathcal{F}_k^{a} = \bigl(M_{t_k}^{\rho(a)} \ge b_k\bigr)
\wedge \bigl(|\mathcal{Q}_{t_k}^{\rho(a)}| < Q_{\rho(a)}^{\max}\bigr).
\end{equation}
For rejection evaluation, let $\widehat T_k^{a}$ be the estimated completion
delay of route $a$ before insertion, and define
\begin{equation}
\mathcal{R}_k^{a}
=
\mathcal{F}_k^{a}
\wedge
\bigl(\widehat T_k^{a}\le D_k\bigr),
\qquad a\in\{0,1\}.
\end{equation}
The closed-loop task outcome is then
\begin{equation}\label{eq:failure_modes}
\chi_k =
\begin{cases}
\mathrm{TP},
& a_k\in\{0,1\},\ \mathcal{F}_k^{a_k},\ T_k\le D_k;\\[1mm]
\mathrm{FP},
& a_k\in\{0,1\},\
  \bigl(\neg\mathcal{F}_k^{a_k}
  \vee(\mathcal{F}_k^{a_k}\wedge T_k>D_k)\bigr);\\[1mm]
\mathrm{TN},
& a_k=2,\ \neg\bigl(\mathcal{R}_k^{0}\vee\mathcal{R}_k^{1}\bigr);\\[1mm]
\mathrm{FN},
& a_k=2,\ \mathcal{R}_k^{0}\vee\mathcal{R}_k^{1}.
\end{cases}
\end{equation}

These constraints define resource feasibility and task-outcome labels for the
long-term net-utility objective. The labels distinguish correct from erroneous
rejections: Rejecting a task is a true negative if no route can complete it by
its deadline and a false negative otherwise. Accepting a task that cannot enter
a queue or complete on time is a false positive. The labels therefore penalize
both blind acceptance under overload and unnecessary rejection.

The objective~\eqref{eq:overall_obj} couples the two event-triggered decisions.
The EN acts at task arrivals and generates $U_k$, whereas the SN selects
$(\sigma_n,g_n)$ at its decision epochs and incurs $C_n$ only when
$\sigma_n=1$. Because one semantic update may affect multiple subsequent
tasks, the update and task-decision subproblems are not independent. We optimize their long-term time average jointly, accounting for task
utility, state staleness, and update cost.

Problem~\eqref{eq:overall_obj} defines the system-level objective, but the
implementation does not combine both event streams in one centralized Bellman
recursion. Instead, CoSMO derives agent-specific returns from the same realized
utility stream. The EN estimates the terminal utility of each task action,
whereas the SN estimates a semi-Markov return over consecutive update intervals.
This decomposition matches the times at which each agent's consequences become
observable while retaining reinforcement learning at both agents.

\section{CoSMO: Cooperative Semantic-Aware Updating and Offloading}
\label{sec:algorithm}

\begin{figure*}[htbp]
\centering
\includegraphics[width=\textwidth]{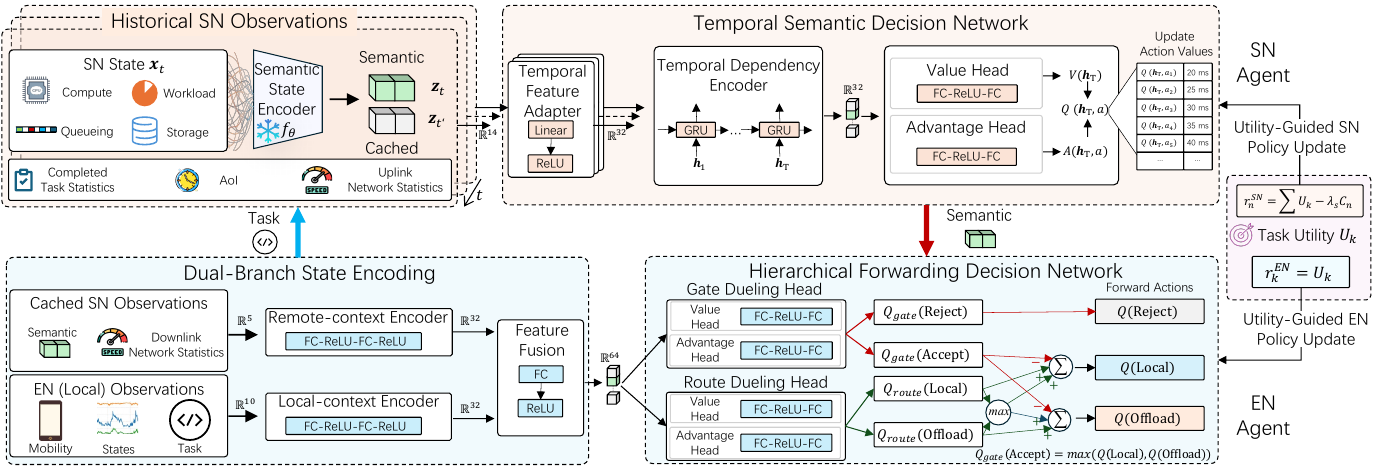}
\caption{Overview of the \textsc{CoSMO} architecture and information flow.
SN information flows from Historical SN Observations through the Semantic State
Encoder, which compresses service states into status semantics, to the Temporal
Semantic Decision Network. Its Temporal Feature Adapter projects observation
features, and its Temporal Dependency Encoder models temporal dependencies; the
Value Head estimates state value and the Advantage Head estimates action
advantages to select the next update interval. EN
information flows from local EN observations and cached SN observations through
Dual-Branch State Encoding, which separately encodes local task/service context
and remote semantic/link context. Feature Fusion combines the two
representations, and the Hierarchical Forwarding Decision Network first
determines acceptance through its gate and then selects local execution or SN
offloading through its route head. The Utility-Guided EN Policy Update and
Utility-Guided SN Policy Update coordinate both agents through the shared task utility.}
\label{fig:arch}
\end{figure*}

\subsection{Architecture Overview}

To solve the coupled update-and-offloading problem, we design \textsc{CoSMO}, a
two-agent event-driven RL framework. The upper branch of
\figurename~\ref{fig:arch} implements the information flow at the SN. The
frozen \emph{encoder} $f_{\theta}(\cdot)$ maps the raw \emph{SN state} into a
compact status semantic. This semantic, its cached copy, AoI, and link/task
statistics form the \emph{SN observation}. A \emph{temporal feature adapter}
then feeds the observation to the \emph{temporal semantic decision network},
which produces \emph{update action values} for each paired action
$(\sigma_n,g_n)$.

The lower branch implements the task flow through \emph{dual-branch state
encoding}. The \emph{remote-context encoder} processes the \emph{cached SN
observation}, while the \emph{local-context encoder} processes the \emph{EN
local observation}. This separation encodes remote-semantic and local-task
features with different dimensions, timescales, and statistical meanings before
fusion. The \emph{feature fusion} module combines both outputs and passes them
to the \emph{hierarchical forwarding decision network}. Its gate and route
dueling heads produce the final \emph{forward actions}: Rejection, local
execution, or offloading.

The \emph{utility-guided learning} branch coordinates the agents without a
centralized execution policy. Its \emph{reward mapping coordinator} maps the
common \emph{task utility} stream $U_k$ to EN per-task rewards and SN interval
rewards. The agents retain separate replay streams and update through the
\emph{EN loss branch} and \emph{SN loss branch}, while the encoder remains
frozen.

\subsection{Raw-State Semantic Encoder}

We implement the encoder $f_{\theta}(\cdot)$
in~\eqref{eq:semantic-encoder} as a lightweight multilayer perceptron (MLP).
The four-dimensional SN raw state
$\mathbf{x}_t^{\mathrm{SN}}$
is first rescaled by a fixed per-dimension normalizer
and then encoded by
\begin{align}
\mathbf{h}_t^{(1)} 
&= \mathrm{ReLU}\!\left(W_1 \mathbf{x}_t^{\mathrm{SN}} + \mathbf{b}_1\right); \\
\mathbf{h}_t^{(2)} 
&= \mathrm{ReLU}\!\left(W_2 \mathbf{h}_t^{(1)} + \mathbf{b}_2\right); \\
\mathbf{z}_t 
&= W_3 \mathbf{h}_t^{(2)} + \mathbf{b}_3,
\end{align}
with parameters $\theta=\{W_1,\mathbf{b}_1,W_2,\mathbf{b}_2,W_3,\mathbf{b}_3\}$.

The output $\mathbf{z}_t$ serves two purposes. It is the remote status semantic
sent to and cached at the EN, and it forms part of the SN scheduler input. We
use a lightweight MLP to map the multidimensional, physically interpretable raw
state into a compact unified representation at low inference cost.

Before utility-coordinated two-agent RL training, we pretrain
$f_{\theta}(\cdot)$ independently with an interval-conditioned forward-dynamics
prediction loss. The selected update interval $g_n$ conditions the predicted
service-state evolution. A lightweight prediction head on
$\mathbf{z}_{u_n}$ computes
\begin{equation}\label{eq:transition_predictor}
\widehat{\Delta \mathbf{x}}_{u_n}
= g_{\psi} \left(
    \left[
        \mathbf{z}_{u_n}, \frac{g_n}{g_{\max}}
    \right]
\right),
\end{equation}
where $g_n$ is the selected SN decision interval, $g_{\max}$ is the maximum
candidate interval, and $\widehat{\Delta \mathbf{x}}_{u_n}$ is the predicted
normalized state increment. The interval-conditioned forward-dynamics loss is
\begin{equation}\label{eq:transition_pretrain}
\mathcal{L}_{\mathrm{pre}}(\theta,\psi)
= \frac{1}{N} \sum_{n=1}^{N} \mathrm{SmoothL1}
\left(
    \widehat{\Delta \mathbf{x}}_{u_n},
    \frac{\mathbf{x}_{u_{n+1}}^{\mathrm{SN}} - \mathbf{x}_{u_n}^{\mathrm{SN}}}{\mathbf{s}}
\right),
\end{equation}
where $\mathbf{s}$ is the per-dimension normalization scale of the raw SN state.
This loss governs the state change associated with the current service state
and selected interval. It avoids contrastive negative samples, which can be
ambiguous when service states repeat.

The learned status semantic supports downstream control rather than raw-state
reconstruction. The prediction loss $\mathcal{L}_{\mathrm{pre}}$ therefore
shapes $\mathbf{z}_t$ to capture dynamics without reconstructing
$\mathbf{x}_t^{\mathrm{SN}}$. After pretraining, the encoder is frozen and
provides a fixed control semantic while only the SN update scheduler and EN
offloading agent are updated.

\begin{figure}[htbp]
\centering
\includegraphics[width=0.96\linewidth]{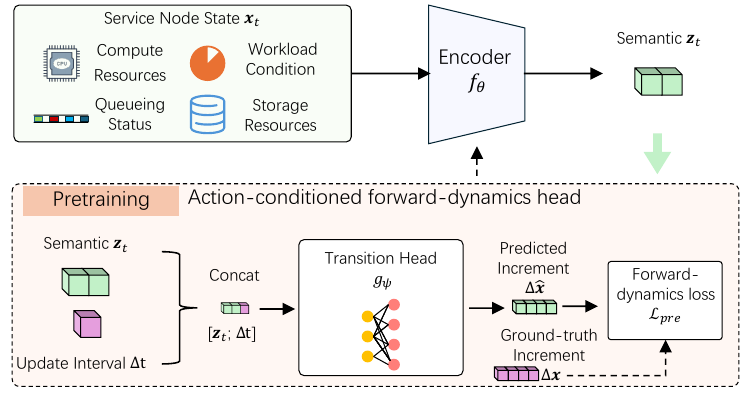}
\caption{Semantic encoder pretraining. The normalized SN raw state is mapped by $f_{\theta}$ to the control
semantic. Together with the normalized update interval, this semantic is fed to $g_{\psi}$ to predict the normalized
state increment for training with $\mathcal{L}_{\mathrm{pre}}$.}
\label{fig:encoder}
\end{figure}

\figurename~\ref{fig:encoder} illustrates the encoder workflow. The encoder
$f_{\theta}$ maps the normalized SN raw state to the control semantic
$\mathbf{z}_t$. At runtime, the SN transmits $\mathbf{z}_t$ as the cached
status and also supplies it to the update scheduler. The interval-conditioned
head $g_{\psi}$ reads $\mathbf{z}_t$ and the selected decision interval only
during the independent pretraining stage in~\eqref{eq:transition_pretrain}.
Then, this auxiliary head is discarded and the encoder is frozen, yielding
a fixed representation shared by the updating and offloading policies.

\subsection{Temporal Semantic Decision Network}

The SN invokes its update scheduler at decision epoch $u_n$. The scheduler
chooses whether to transmit the current semantic and how long to wait before
the next decision. The candidate interval set is
\begin{equation}
\mathcal{G} = \bigl\{
    g^{(1)}, g^{(2)}, \ldots, g^{(L)}
\bigr\},
\end{equation}
with
\begin{equation}
g_{\min} \le g^{(1)} < g^{(2)} < \cdots < g^{(L)} \le g_{\max}.
\end{equation}
With the no-send option enabled, the discrete SN action set is
\begin{equation}\label{eq:sn_action_set_impl}
\mathcal{A}^{\mathrm{SN}}
= \bigl\{
    (\sigma,g^{(\ell)}):
    \sigma\in\{0,1\},\ \ell=1,\ldots,L
\bigr\},
\end{equation}
where $\sigma=1$ denotes sending the current semantic and $\sigma=0$ denotes
skipping the current transmission. If no-send is disabled,
$\mathcal{A}^{\mathrm{SN}}$ reduces to the always-send subset
$\{(1,g^{(\ell)}):\ell=1,\ldots,L\}$.

From $\mathcal{I}_n^{\mathrm{SN}}$ in~\eqref{eq:sn_info_set}, we form the SN
observation vector as
\begin{equation}\label{eq:sn_obs}
\begin{aligned}
\mathbf{o}_n^{\mathrm{SN}}
&= \Gamma^{\mathrm{SN}}\!\bigl(\mathcal{I}_n^{\mathrm{SN}}\bigr)\\
&= \bigl[
    \mathbf{z}_{u_n},\,
    \mathbf{z}^c_{\nu_n},\,
    \widetilde A_{u_n},\,
    \psi_H\!\bigl(\mathcal{H}_n^{\mathrm{comp}}\bigr),\\[-1mm]
    &\hspace{16mm}
    \psi_{\ell}^{\mathrm{SN}}\!\bigl(\ell_{u_n}^{\mathrm{SN}}\bigr)
\bigr],
\end{aligned}
\end{equation}
where $\Gamma^{\mathrm{SN}}(\cdot)$ is the feature construction,
$\mathbf{z}_{u_n}=f_{\theta}(\mathbf{x}_{u_n}^{\mathrm{SN}})$ is the current
raw state's semantic, and $\mathbf{z}^c_{\nu_n}$ is the semantic cached at the
EN before the new SN action. The term $\widetilde A_{u_n}$ is the normalized
age of the cached semantic.
$\psi_H(\mathcal{H}_n^{\mathrm{comp}})$ comprises normalized statistics over
offloaded tasks completed within the most recent window, including completion
delay and success/failure rates.
$\psi_{\ell}^{\mathrm{SN}}(\ell_{u_n}^{\mathrm{SN}})$ comprises the short-term
mean rate, rate standard deviation, and rate trend of the SN-to-EN update link.
Correspondingly,
$\psi_{\ell}^{\mathrm{EN}}(\ell_{t_k}^{\mathrm{EN}})$ below uses the same kind of
statistics for the EN-to-SN link.

An SN update affects the objective through tasks served during the subsequent
interval rather than immediately. We therefore project each SN observation into
a feature vector and feed the resulting sequence into the GRU. Let
\begin{equation}
\mathbf{e}_n^{\mathrm{SN}}
= \phi_{\omega_p}^{\mathrm{SN}}\left(\mathbf{o}_n^{\mathrm{SN}}\right)
\end{equation}
denote the projected SN feature, and let $\mathbf{h}_n^{\mathrm{SN}}$ be the
hidden state for the $n$-th decision:
\begin{equation}\label{eq:sn_gru}
\mathbf{h}_n^{\mathrm{SN}}
= \mathrm{GRU}_{\omega_g} \left(
    \mathbf{e}_n^{\mathrm{SN}}, \mathbf{h}_{n-1}^{\mathrm{SN}}
\right),
\end{equation}
with projection parameters $\omega_p$ and GRU parameters $\omega_g$.

We apply a dueling Double DQN to $\mathbf{h}_n^{\mathrm{SN}}$, using a
state-value branch and an advantage branch:
\begin{align}
V_{\omega_v}^{\mathrm{SN}} \left(\mathbf{h}_n^{\mathrm{SN}}\right),
\qquad
A_{\omega_a}^{\mathrm{SN}} \left(\mathbf{h}_n^{\mathrm{SN}}, a\right),
\quad a\in\mathcal{A}^{\mathrm{SN}},
\end{align}
and combine them into the SN action-value function
\begin{equation}\label{eq:sn_dueling_q}
\begin{split}
Q_{\omega}^{\mathrm{SN}}\!\left(\mathbf{h}_n^{\mathrm{SN}}, a\right)
= & V_{\omega_v}^{\mathrm{SN}}\!\left(\mathbf{h}_n^{\mathrm{SN}}\right)
    + A_{\omega_a}^{\mathrm{SN}}\!\left(\mathbf{h}_n^{\mathrm{SN}}, a\right)\\
  & - \frac{1}{|\mathcal{A}^{\mathrm{SN}}|} \sum_{\tilde a\in\mathcal{A}^{\mathrm{SN}}}
A_{\omega_a}^{\mathrm{SN}}\!\left(\mathbf{h}_n^{\mathrm{SN}}, \tilde a\right),
\end{split}
\end{equation}
with $\omega=\{\omega_p,\omega_g,\omega_v,\omega_a\}$. The network takes
$\mathbf{o}_n^{\mathrm{SN}}$ and the previous hidden state
$\mathbf{h}_{n-1}^{\mathrm{SN}}$. It outputs
$[Q_{\omega}^{\mathrm{SN}}(\mathbf{h}_n^{\mathrm{SN}},a)]_{a\in\mathcal{A}^{\mathrm{SN}}}
\in\mathbb{R}^{18}$ over the SN action set, yielding the decision
\begin{equation}\label{eq:sn_policy}
a_n^{\mathrm{SN}} = 
    \arg\max_{a\in\mathcal{A}^{\mathrm{SN}}}     Q_{\omega}^{\mathrm{SN}}\!\left(\mathbf{h}_n^{\mathrm{SN}}, a\right),
    \qquad
    a_n^{\mathrm{SN}}=(\sigma_n,g_n).
\end{equation}

To align the SN action values with the overall objective, the return for the
$n$th decision includes all tasks arriving in its selected interval:
\begin{equation}
\mathcal{K}_n = \{k:u_n\le t_k<u_{n+1}\}.
\end{equation}
The base interval reward is
\begin{equation}\label{eq:sn_reward_new}
r_{n,0}^{\mathrm{SN}} = 
\beta_{\mathrm{SN}} \sum_{k\in\mathcal{K}_n} U_k - \lambda_s C_n,
\end{equation}
This reward sums the task utilities over the following interval, optionally
scaled by $\beta_{\mathrm{SN}}$ for ablations, and subtracts the explicit
semantic-update cost. In the main objective, $\beta_{\mathrm{SN}}=1$. Since
$C_n=\sigma_n c_s d_z$, the update cost is charged only for $\sigma_n=1$.
When no task arrives in the interval, the task-utility sum is zero. The trainer
also supports an optional AP-value stale-loss shaping term in ablations; when
its configured weight is zero, the reward used by the SN replay is exactly
$r_n^{\mathrm{SN}}=r_{n,0}^{\mathrm{SN}}$.

Double DQN separates next-action \emph{selection} and \emph{evaluation} across
two networks to reduce Q-value overestimation. The online network computes the
next hidden state $\mathbf{h}_{n+1}^{\mathrm{SN}}$ and selects the greedy action:
\begin{equation}\label{eq:sn_next_greedy}
a_{n+1}^{\star} = \arg\max_{a'\in\mathcal{A}^{\mathrm{SN}}} Q_{\omega}^{\mathrm{SN}} \left(\mathbf{h}_{n+1}^{\mathrm{SN}}, a'\right),
\end{equation}
The target network computes $\bar{\mathbf{h}}_{n+1}^{\mathrm{SN}}$ and evaluates
the selected action. Because consecutive SN decisions are separated by
heterogeneous intervals $g_n$, a semi-Markov discount assigns greater
discounts to longer waits. The implementation uses the normalized reward
\(\widetilde r_n^{\mathrm{SN}}=r_n^{\mathrm{SN}}/s_r^{\mathrm{SN}}\), where
$s_r^{\mathrm{SN}}$ takes the base task value $V$ by default. With terminal indicator
$d_n$, the SN target value is
\begin{equation}\label{eq:sn_target_new}
y_n^{\mathrm{SN}} = \widetilde r_n^{\mathrm{SN}}
+
(1-d_n) \gamma_{\mathrm{SN}}^{g_n/g_0} Q_{\omega^-}^{\mathrm{SN}}
\left(
    \bar{\mathbf{h}}_{n+1}^{\mathrm{SN}}, a_{n+1}^{\star}
\right),
\end{equation}
with target parameters $\omega^-$, discount base $\gamma_{\mathrm{SN}}\in(0,1)$, and
reference interval $g_0=g^{(1)}$ (the shortest candidate in $\mathcal{G}$). The
exponent $g_n/g_0$ measures the elapsed interval in units of $g_0$. The SN training
loss is
\begin{equation}\label{eq:sn_loss_new}
\mathcal{L}_{\mathrm{SN}}(\omega) = \mathbb{E}
    \Bigl[
        \mathrm{SmoothL1} \bigl(
            Q_{\omega}^{\mathrm{SN}}(\mathbf{h}_n^{\mathrm{SN}}, a_n^{\mathrm{SN}}),
            y_n^{\mathrm{SN}}
        \bigr)
    \Bigr],
\end{equation}
where $\mathrm{SmoothL1}(\cdot,\cdot)$ is the Huber loss ($\beta=1$). It is less
sensitive than squared error to the heavier-tailed SN targets induced by
interval-level returns and sequence bootstrap. The replay buffer samples
contiguous SN transition sequences, and the GRU is unrolled over each sequence
during the update. During utility-coordinated training, the encoder
$f_{\theta}(\cdot)$ is frozen. Thus,
$\mathbf{z}_{u_n}=f_{\theta}(\mathbf{x}_{u_n}^{\mathrm{SN}})$ is recomputed only to
construct the current SN observation, while $\mathbf{z}^c_{\nu_n}$ is taken
directly from the cached EN copy stored in replay and bypasses re-encoding.

\subsection{EN Selective-Offloading Agent}

At each task arrival, the EN selective-offloading agent selects one action from
$\mathcal{A}^{\mathrm{EN}}=\{0,1,2\}$: Local execution, offloading to the SN, or
active rejection. As shown in the lower branch of
\figurename~\ref{fig:arch}, the agent encodes the directly observed local
context separately from the cached remote SN context. A hierarchical
gate--route network then processes the fused representation. This design
reflects the different reliability and timescales of the two information
sources. Local task, access, and EN service features are current and directly
observed, whereas the remote semantic is compact, delayed, and affected by the
update schedule and inter-node link. The gate makes the accept/reject decision
under resource and deadline risk. Conditional on acceptance, the route head
selects local execution or SN offloading. This structure preserves local feasibility cues, learns how much to trust stale remote semantics,
and reduces interference between rejection and route selection.

\subsubsection{Dual-Branch State Encoding}

From $\mathcal{I}_k^{\mathrm{EN}}$ in~\eqref{eq:ap_info_set}, we construct the
local and remote inputs:
\begin{align}
\mathbf{o}_{k,\mathrm{loc}}^{\mathrm{EN}} &=
    \Bigl[
        \mathbf{m}_k,\,
        \mathbf{x}_{t_k}^{\mathrm{EN}},\, (b_k,c_k,D_k)
    \Bigr];
\label{eq:ap_local_obs}\\
\mathbf{o}_{k,\mathrm{rem}}^{\mathrm{EN}} &=
    \Bigl[
        \mathbf{z}^c_{u_{n(k)}},\,
        \psi_{\ell}^{\mathrm{EN}}
         \bigl(\ell_{t_k}^{\mathrm{EN}}\bigr)
    \Bigr],
\label{eq:ap_remote_obs}
\end{align}
where $\mathbf{o}_{k,\mathrm{loc}}^{\mathrm{EN}}$ collects the directly
observed access, task, and EN service state at arrival. The remote observation
$\mathbf{o}_{k,\mathrm{rem}}^{\mathrm{EN}}$ contains the cached SN status
semantic and inter-node link information. In particular,
$\psi_{\ell}^{\mathrm{EN}}(\ell_{t_k}^{\mathrm{EN}})$ comprises the short-term
mean rate, rate standard deviation, and rate trend of the EN-to-SN link.

Separate two-layer MLPs encode the two inputs:
\begin{align}
\mathbf{h}_{k,\mathrm{loc}}^{\mathrm{EN}}
& =
f_{\varphi_{\ell}}^{\mathrm{loc}}
\left(
    \mathbf{o}_{k,\mathrm{loc}}^{\mathrm{EN}}
\right),\\
\mathbf{h}_{k,\mathrm{rem}}^{\mathrm{EN}}
& = f_{\varphi_{r}}^{\mathrm{rem}}
\left(
    \mathbf{o}_{k,\mathrm{rem}}^{\mathrm{EN}}
\right),
\end{align}
where $\varphi_{\ell}$ and $\varphi_r$ denote the two branch parameter sets.
The resulting features are fused into the EN decision representation
\begin{equation}
\label{eq:ap_fusion}
\mathbf{h}_k^{\mathrm{EN}} = \mathrm{ReLU} \left(
    W_f \left[
        \mathbf{h}_{k,\mathrm{loc}}^{\mathrm{EN}},\,
        \mathbf{h}_{k,\mathrm{rem}}^{\mathrm{EN}}
    \right]
    + \mathbf{b}_f
\right),
\end{equation}
where $[\,\cdot,\cdot\,]$ denotes concatenation, and $W_f$ and$\mathbf{b}_f$ are the fusion parameters.

\subsubsection{Hierarchical Forwarding Decision Network}

Given the fused representation $\mathbf{h}_k^{\mathrm{EN}}$, the hierarchical
forwarding decision network separates accept/reject from route selection. The
gate head scores $\mathcal{B}=\{\mathrm{reject},\mathrm{accept}\}$, whereas the route head scores $\mathcal{R}=\{0,1\}$, corresponding to local execution and SN offloading.
Each head adopts a dueling value--advantage decomposition:
\begin{align}
V_{\varphi_{gv}}^{\mathrm{gate}} \left(\mathbf{h}_k^{\mathrm{EN}}\right), \quad A_{\varphi_{ga}}^{\mathrm{gate}} \left(\mathbf{h}_k^{\mathrm{EN}},b\right), \quad b\in\mathcal{B};\\
V_{\varphi_{\rho v}}^{\mathrm{route}} \left(\mathbf{h}_k^{\mathrm{EN}}\right), \quad A_{\varphi_{\rho a}}^{\mathrm{route}} \left(\mathbf{h}_k^{\mathrm{EN}},r\right), \quad r\in\mathcal{R}.
\end{align}
The corresponding dueling scores are
\begin{align}
Q_{\varphi}^{\mathrm{gate}} \left(\mathbf{h}_k^{\mathrm{EN}},b\right)
& = V_{\varphi_{gv}}^{\mathrm{gate}} \left(\mathbf{h}_k^{\mathrm{EN}}\right) +
A_{\varphi_{ga}}^{\mathrm{gate}} \!\left(\mathbf{h}_k^{\mathrm{EN}},b\right)
\nonumber\\ 
    &\quad - \frac{1}{|\mathcal{B}|} \sum_{\tilde b\in\mathcal{B}} A_{\varphi_{ga}}^{\mathrm{gate}} \!\left(\mathbf{h}_k^{\mathrm{EN}},\tilde b\right);
\label{eq:ap_gate_dueling_q}\\
Q_{\varphi}^{\mathrm{route}} \left(\mathbf{h}_k^{\mathrm{EN}},r\right)
& =
    V_{\varphi_{\rho v}}^{\mathrm{route}}
    \left(\mathbf{h}_k^{\mathrm{EN}}\right)
    +
    A_{\varphi_{\rho a}}^{\mathrm{route}}
    \left(\mathbf{h}_k^{\mathrm{EN}},r\right)
    \nonumber\\
    &\quad - \frac{1}{|\mathcal{R}|}
    \sum_{\tilde r\in\mathcal{R}}
    A_{\varphi_{\rho a}}^{\mathrm{route}} \left(\mathbf{h}_k^{\mathrm{EN}},\tilde r\right).
\label{eq:ap_route_dueling_q}
\end{align}

We combine the two heads through a hierarchy-consistent action-value
reconstruction. The reject action retains the gate's reject value. Each
accepted action combines the gate's accept value with its max-normalized route
score:
\begin{equation}
\label{eq:ap_dueling_q}
Q_{\varphi}^{\mathrm{EN}} \left(\mathbf{h}_k^{\mathrm{EN}},a\right) =
\begin{cases}
\begin{aligned}
& Q_{\varphi}^{\mathrm{gate}} \left(\mathbf{h}_k^{\mathrm{EN}},\mathrm{accept}\right)\\
    &\quad + Q_{\varphi}^{\mathrm{route}} \left(\mathbf{h}_k^{\mathrm{EN}},a\right)\\
    &\quad - \max_{\tilde r\in\mathcal{R}} Q_{\varphi}^{\mathrm{route}} \left(\mathbf{h}_k^{\mathrm{EN}},\tilde r\right)
\end{aligned},
& a\in\{0,1\};\\[1mm]

Q_{\varphi}^{\mathrm{gate}} \left(\mathbf{h}_k^{\mathrm{EN}},\mathrm{reject}\right),
& a=2.
\end{cases}
\end{equation}
Under this reconstruction, the gate head determines the value of the accepted
subtree. The route head represents only the relative preference between local
execution and offloading within that subtree. Specifically,
\begin{equation}
\label{eq:ap_hierarchical_consistency}
\max_{a\in\{0,1\}} Q_{\varphi}^{\mathrm{EN}} \left(\mathbf{h}_k^{\mathrm{EN}},a\right)
=
Q_{\varphi}^{\mathrm{gate}} \left( \mathbf{h}_k^{\mathrm{EN}},\mathrm{accept} \right),
\end{equation}
and
\begin{align}
&
Q_{\varphi}^{\mathrm{EN}} \left(\mathbf{h}_k^{\mathrm{EN}},0\right)
    - Q_{\varphi}^{\mathrm{EN}}\left(\mathbf{h}_k^{\mathrm{EN}},1\right)
\nonumber\\
&\qquad = 
Q_{\varphi}^{\mathrm{route}} \left(\mathbf{h}_k^{\mathrm{EN}},0\right)
    - Q_{\varphi}^{\mathrm{route}}\left(\mathbf{h}_k^{\mathrm{EN}},1\right).
\label{eq:ap_route_preference_consistency}
\end{align}
Equation~\eqref{eq:ap_hierarchical_consistency} assigns the gate's accept value
to the best accepted action. Equation~\eqref{eq:ap_route_preference_consistency}
preserves the local/offload preference learned by the route head. In this sense, gate governs the top-level comparison, while the route head determines the
conditional choice within the accepted branch.

We define
\begin{align}
q_{k}^{\mathrm{rej}}
& = Q_{\varphi}^{\mathrm{gate}}\left(
    \mathbf{h}_k^{\mathrm{EN}},\mathrm{reject}
\right),
\label{eq:ap_reject_score}\\
q_{k}^{\mathrm{acc}}
& = Q_{\varphi}^{\mathrm{gate}}
    \left(
        \mathbf{h}_k^{\mathrm{EN}},\mathrm{accept}
    \right).
\label{eq:ap_accept_score}
\end{align}
The gate-first decoder selects
\begin{equation}
\label{eq:ap_policy_new}
a_k = \begin{cases}
    2,
    & q_{k}^{\mathrm{rej}}\ge q_{k}^{\mathrm{acc}},\\[1mm]
    \displaystyle
    \arg\max_{r\in\mathcal{R}} Q_{\varphi}^{\mathrm{route}}
        \left(\mathbf{h}_k^{\mathrm{EN}},r\right),
    & q_{k}^{\mathrm{rej}}<q_{k}^{\mathrm{acc}}.
\end{cases}
\end{equation}
The first stage makes the accept/reject decision. The second selects an
execution route only for the accepted branch. The training-only gate and route
auxiliary losses introduced below regularize these conditional decisions but
are not used during execution.

To align EN learning with the system objective, each task-level transition
receives the realized net task utility
\begin{equation}
\label{eq:ap_reward_new}
r_k^{\mathrm{EN}}=U_k.
\end{equation}
The value of $U_k$ may be unavailable at task arrival because it depends on the
realized delay $T_k$ and deadline outcome. For active rejection ($a_k=2$), the
utility is immediately determined as a true-negative reward or false-negative
penalty, according to~\eqref{eq:task_utility} and~\eqref{eq:failure_modes}, respectively. The
false-positive utility is also available before execution if a local or offload
attempt fails the selected-node acceptance check. Otherwise, for an accepted
task, the EN first stores the transition prefix
\begin{equation}
\tilde e_{k,\mathrm{pre}}^{\mathrm{EN}} =
\left(
    \mathbf{o}_{k,\mathrm{loc}}^{\mathrm{EN}},
    \mathbf{o}_{k,\mathrm{rem}}^{\mathrm{EN}},
    a_k
\right).
\end{equation}
After the task reaches a terminal outcome, the EN fills in
$r_k^{\mathrm{EN}}=U_k$ and writes the completed transition to the replay
buffer.

During an EN optimizer update, the stored utility is normalized as
\begin{equation}
\widetilde{U}_k = \frac{U_k}{s_r^{\mathrm{EN}}},
\qquad
s_r^{\mathrm{EN}}=V.
\end{equation}
This positive scaling preserves the action ordering while controlling the
numerical range of targets across the four outcome classes.

The EN learns the expected terminal utility of each hierarchical action from
task-level environment feedback rather than minimizing only instantaneous
transmission delay. Its value estimate is conditioned on the current access
state, local load, cached remote semantic, and task timeliness requirement.
Under overload, the learned policy may reject tasks that are unlikely to
complete on time, avoiding their transmission, queueing, and compute costs.

Each EN replay sample is completed only after the corresponding task outcome is
known. We model the finalized task as a terminal EN transition. The general
Bellman target therefore reduces to the normalized realized utility because the
next-state bootstrap term vanishes:
\begin{equation}
\label{eq:ap_target_new}
y_k^{\mathrm{EN}} = \frac{r_k^{\mathrm{EN}}}{s_r^{\mathrm{EN}}}
        + \gamma_{\mathrm{EN}} \left(1-d_k^{\mathrm{EN}}\right)
        \widehat{Q}_{k+1}^{\mathrm{EN}} 
    = \widetilde{U}_k,
    \qquad
    d_k^{\mathrm{EN}}=1,
\end{equation}
where $\widehat{Q}_{k+1}^{\mathrm{EN}}$ denotes a generic next-state bootstrap
value, which is not evaluated for the EN's task-terminal transitions. The
resulting temporal-difference loss is
\begin{equation}
\label{eq:ap_loss_new}
\mathcal{L}_{\mathrm{EN}}^{\mathrm{TD}}(\varphi) = \mathbb{E}
\left[
    \mathrm{SmoothL1} \left(
        Q_{\varphi}^{\mathrm{EN}} \left(\mathbf{h}_k^{\mathrm{EN}},a_k\right),  y_k^{\mathrm{EN}}
    \right)
\right].
\end{equation}
Accordingly, the EN action-value function estimates
\begin{equation}
Q_{\varphi}^{\mathrm{EN}} \left(\mathbf{h}_k^{\mathrm{EN}},a_k\right)
\approx \mathbb{E} \left[
    \widetilde{U}_k \mid \mathbf{h}_k^{\mathrm{EN}},a_k
\right].
\end{equation}
The EN remains an off-policy, value-based RL agent trained through
exploration, environment feedback, and experience replay. Its task-terminal
formulation does not require inter-task bootstrapping. By contrast, the SN
scheduler retains a semi-Markov Double DQN target across consecutive update
intervals.

The remote semantic in $\mathbf{o}_{k,\mathrm{rem}}^{\mathrm{EN}}$ is the
historical copy transmitted by the SN and cached during interaction. The EN replay
stores this copy and its remote-context features directly, rather than
regenerating them from the instantaneous raw SN state during training. This
preserves consistency with decentralized online execution. The EN loss updates only the offloading-agent parameters $\varphi$; the encoder
parameters $\theta$ are pretrained and remain frozen throughout
utility-coordinated training.

\subsection{Utility-Coordinated Training Algorithm}

The two trainable agents share an event-driven environment and are coordinated
through the realized task-utility stream in
\figurename~\ref{fig:arch}. The encoder $f_{\theta}(\cdot)$ is pretrained and
then frozen. Utility-coordinated training updates only the SN scheduler
$Q_{\omega}^{\mathrm{SN}}(\cdot,\cdot)$ and EN selective-offloading agent
$Q_{\varphi}^{\mathrm{EN}}(\cdot,\cdot)$. The reward coordinator maps each
finalized task utility to the EN reward $r_k^{\mathrm{EN}}$. It also aggregates
the utilities realized over $[u_n,u_{n+1})$ into the SN reward
$r_n^{\mathrm{SN}}$, together with the optional normalization and shaping terms
defined above.

The agents use separate observations, replay buffers, and value losses. The EN
learns a task-terminal action-value function from finalized outcomes, whereas
the SN learns a recurrent semi-Markov action-value function across update
intervals. The common task-utility stream coordinates these agents without a
centralized critic. Optional oracle-derived gate--route labels regularize EN
training but are unavailable during decentralized execution.
Algorithm~\ref{alg:ctde_training} summarizes the complete procedure.

\begin{algorithm}[!t]
\caption{Utility-coordinated training of semantic updating and selective
offloading.}
\label{alg:ctde_training}
\footnotesize
\begin{algorithmic}[1]
\REQUIRE Shared environment $\mathcal{E}$, frozen pretrained encoder
$f_{\theta}$, SN online and target parameters $\omega$ and $\omega^{-}$,
EN parameters $\varphi$, and replay buffers
$\mathcal{D}_{\mathrm{SN}}$ and $\mathcal{D}_{\mathrm{EN}}$
\FOR{each training episode}
    \STATE Reset $\mathcal{E}$ and clear the EN cached semantic and SN GRU
    state
    \label{alg:line-reset}
    \WHILE{the episode is unfinished}
        \IF{an SN decision event $u_n$ occurs}
            \STATE Encode
            $\mathbf{z}_{u_n}
            =f_{\theta}(\mathbf{x}_{u_n}^{\mathrm{SN}})$,
            build $\mathbf{o}_n^{\mathrm{SN}}$, select
            $a_n^{\mathrm{SN}}=(\sigma_n,g_n)$ using
            \eqref{eq:sn_policy}, transmit $\mathbf{z}_{u_n}$ if
            $\sigma_n=1$, and schedule the next decision at
            $u_{n+1}=u_n+g_n$
            \label{alg:line-sn-decision}
        \ENDIF
        \IF{a transmitted SN semantic reaches the EN}
            \STATE Update the EN semantic cache with the delivered semantic
            and its source timestamp
        \ENDIF
        \IF{a task arrival $t_k$ occurs}
            \STATE Build the EN local and cached-remote observations, select
            $a_k$ using \eqref{eq:ap_policy_new}, and store an EN transition
            prefix if its utility is not yet known
            \label{alg:line-en-arrival}
        \ENDIF
        \IF{a task outcome is determined}
            \STATE Finalize $U_k$ using \eqref{eq:task_utility}, assign
            $r_k^{\mathrm{EN}}=U_k$, complete the EN transition, and write it
            to $\mathcal{D}_{\mathrm{EN}}$
            \label{alg:line-en-reward}
        \ENDIF
        \IF{the update interval $[u_n,u_{n+1})$ ends}
            \STATE Compute $r_n^{\mathrm{SN}}$ using
            \eqref{eq:sn_reward_new} and write the corresponding SN sequence
            sample to $\mathcal{D}_{\mathrm{SN}}$
            \label{alg:line-sn-reward}
        \ENDIF
        \STATE At the configured frequencies, update $\varphi$ from
        $\mathcal{D}_{\mathrm{EN}}$ using the task-terminal EN loss and
        update $\omega$ from $\mathcal{D}_{\mathrm{SN}}$ using the
        semi-Markov Double DQN loss, with $f_{\theta}$ frozen; synchronize
        $\omega^{-}$ from $\omega$ when scheduled
        \label{alg:line-param-update}
    \ENDWHILE
\ENDFOR
\end{algorithmic}
\end{algorithm}

Algorithm~\ref{alg:ctde_training} aligns the training loop with the event-driven
execution interface. Line~\ref{alg:line-reset} initializes the episode.
Line~\ref{alg:line-sn-decision} selects SN transmission and the next decision
interval, whereas Line~\ref{alg:line-en-arrival} handles the EN task decision.
A transmitted semantic affects the EN only after delivery to the EN cache.
Because the utility of an accepted task is known only at its terminal outcome,
Line~\ref{alg:line-en-reward} completes the delayed EN replay entry with the
realized utility. Line~\ref{alg:line-sn-reward} assigns the interval reward to
the SN sequence sample. Line~\ref{alg:line-param-update} then performs the
agent-specific parameter updates.

The EN updates follow
\eqref{eq:ap_target_new}--\eqref{eq:ap_loss_new}. Since each completed EN
transition is task-terminal, its value target contains no next-state bootstrap
term and therefore does not require target-network evaluation. When enabled,
the training-only gate--route auxiliary losses are added to the EN loss but are
not used during execution. These auxiliary terms are implementation-level
regularizers and are specified with the training hyperparameters in
Section~\ref{subsec:agent_impl}.

The SN updates follow
\eqref{eq:sn_target_new}--\eqref{eq:sn_loss_new}. The target parameters
$\omega^{-}$ are delayed copies of the online parameters $\omega$. They
evaluate the next action selected by the online network in the semi-Markov
Double DQN target and are synchronized at the configured frequency. The
training procedure thereby coordinates semantic updating and selective
offloading through a common utility objective while retaining agent-specific
value-learning mechanisms.

During decentralized execution, each agent uses only locally available
observations, without real-time global state or training-only oracle
information. The SN outputs
$a_n^{\mathrm{SN}}=(\sigma_n,g_n)$ from $\mathcal{I}_n^{\mathrm{SN}}$ via
\eqref{eq:sn_obs}, \eqref{eq:sn_gru}, and \eqref{eq:sn_policy}. The EN outputs
$a_k$ from $\mathcal{I}_k^{\mathrm{EN}}$ via \eqref{eq:ap_local_obs}, \eqref{eq:ap_remote_obs}, \eqref{eq:ap_fusion}, and \eqref{eq:ap_policy_new}. Thus, both execution policies are realizable from local information.

\subsection{Complexity Analysis}

Let $C_{\mathrm{EN}}^{\mathrm{tr}}$ denote the EN network's per-sample training
cost. It includes forward and backward propagation through the local and remote
encoders, feature-fusion layer, and hierarchical gate--route heads. Let
$C_{\mathrm{SN}}^{\mathrm{tr}}$ denote the SN scheduler's per-step training
cost. It includes the frozen semantic-encoder forward pass, forward and
backward propagation through the projection, GRU, and dueling head, and SN
target-network evaluation.

For one training episode with $M_{\mathrm{EN}}$ EN optimizer updates,
$M_{\mathrm{SN}}$ SN optimizer updates, EN batch size $B_{\mathrm{EN}}$, SN
batch size $B_{\mathrm{SN}}$, and sampled SN sequence length $\tau+1$, the
neural training cost is
\begin{equation}
\mathcal{O}\!\left(
    M_{\mathrm{EN}}B_{\mathrm{EN}} \bigl(
        C_{\mathrm{EN}}^{\mathrm{tr}}+C_{\mathrm{aux}}
    \bigr) +
    M_{\mathrm{SN}}B_{\mathrm{SN}}(\tau+1)
    C_{\mathrm{SN}}^{\mathrm{tr}}
\right),
\end{equation}
where $C_{\mathrm{aux}}$ denotes the per-sample cost of generating and applying
the training-only oracle-derived gate--route auxiliary supervision. With
bounded EN and SN queues, this additional cost is constant with respect to the
episode length.

Replay memory scales as
\begin{equation}
\mathcal{O}\!\left(
    |\mathcal{D}_{\mathrm{EN}}|d_{\mathrm{EN}} +
    |\mathcal{D}_{\mathrm{SN}}|(\tau+1)d_{\mathrm{SN}}
\right),
\end{equation}
where $d_{\mathrm{EN}}$ and $d_{\mathrm{SN}}$ denote the stored dimensions of
one EN transition and one SN transition step, respectively.

Let $C_{\mathrm{EN}}^{\mathrm{inf}}$ and
$C_{\mathrm{SN}}^{\mathrm{inf}}$ denote the corresponding per-decision
inference costs. At execution, replay sampling, oracle labeling,
target-network evaluation, and backpropagation are removed. Therefore, an
episode with $N_{\mathrm{EN}}$ task decisions and $N_{\mathrm{SN}}$ update
decisions incurs
\begin{equation}
\mathcal{O} \left( N_{\mathrm{EN}}C_{\mathrm{EN}}^{\mathrm{inf}}
            +
            N_{\mathrm{SN}}C_{\mathrm{SN}}^{\mathrm{inf}}
\right),
\end{equation}
which is constant time per decision event for fixed network widths and bounded
action sets.

\section{Experiments}
\label{sec:experiments}

We evaluate the proposed cooperative semantic-update and task-offloading policy
when only stale remote-state information is available. The experiments assess
whether the learned EN--SN policy can (i) make load-aware accept/reject
decisions near and beyond the compute-capacity boundary, (ii) maintain on-time
task completion, and (iii) reduce semantic-update overhead relative to
freshness-driven policies. All methods use the same simulation environment,
workload generation procedure, and measurement protocol.
\subsection{Experiment Settings}

\subsubsection{Simulation Environment}
\label{subsec:env}

The simulation is implemented in the \mbox{ns-3.42} discrete-event network
simulator~\cite{ns3} and connected to the Python learning process through the
ns3-gym OpenGym bridge~\cite{ns3gym}. The topology contains four UEs, one EN,
one forwarding relay, and one SN. The relay supports only two-hop EN--SN
forwarding and provides no compute resources. Table~\ref{tab:env} summarizes
the environment parameters.

\begin{table}[!t]
\renewcommand{\arraystretch}{1.3}
\caption{Simulation environment configuration.}
\label{tab:env}
\centering
\footnotesize
\begin{tabular}{p{2.5cm}p{5.0cm}}
\hline
\bfseries Component & \bfseries Configuration \\
\hline
UE & 4 nodes with straight-line constant-velocity mobility at
  \SI{1}{m/s}; initial EN distances \SIlist{5.2;6.0;6.8;5.2}{m};
  \SI{8}{m} service radius \\
UE--EN access link & IEEE 802.11n at \SI{5}{GHz} with \SI{40}{MHz}
  channel bandwidth; effective throughput ${\sim}\SI{90}{}$--$\SI{100}{Mbps}$ \\
EN--SN backhaul & two-hop point-to-point path via one relay;
  EN$\rightarrow$SN \SI{96}{Mbps}, SN$\rightarrow$EN \SI{192}{Mbps},
  \SI{1}{ms} per hop \\
EN capability & 2 compute slots $\times$ \SI{18000}{cycles/ms}; storage
  \SI{240000}{B}; queue capacity 64 tasks \\
SN capability & 4 compute slots $\times$ \SI{27000}{cycles/ms}; storage
  \SI{480000}{B}; queue capacity 64 tasks \\
Episode protocol & \SI{10}{s} arrival window; \SI{2}{s} measurement warm-up;
  simulation continues until all generated tasks terminate \\
\hline
\end{tabular}
\end{table}

The controlled single-EN/single-SN topology holds the network structure, access
conditions, and backhaul rates fixed across methods. The evaluation
isolates semantic updating, task acceptance, and execution routing from
topology and rate adaptation. Tasks arrive during a \SI{10}{s} window in each
episode, after which the simulation continues until every task reaches a
terminal state. UE mobility determines the remaining connection time.
Packet-transfer time includes serialization, queueing, propagation, and
task-dependent transfer delays.

The backhaul uses the fixed asymmetric directional rates in
Table~\ref{tab:env}. These experiments do not assess adaptation to
backhaul-rate variability or short-term rate trends. For each UE--EN link, the
payload size and measured upload delay determine the effective uplink-rate
estimate, which is then smoothed by an EWMA.
\subsubsection{Agent Implementation}
\label{subsec:agent_impl}

The learned policies are trained and selected using the task-utility parameters
in Table~\ref{tab:reward_training} and~\eqref{eq:task_utility}. The
training-only oracle described in Section~\ref{sec:algorithm} provides the
auxiliary gate and route labels. Table~\ref{tab:hparams} reports the core
architecture, action spaces, and training parameters.

\begin{table}[htbp]
\renewcommand{\arraystretch}{1.15}
\caption{Reward and training configuration.}
\label{tab:reward_training}
\centering
\scriptsize
\setlength{\tabcolsep}{3pt}
\begin{tabular}{@{}p{2.45cm}p{4.85cm}@{}}
\hline
\bfseries Category & \bfseries Parameter settings \\
\hline
Task-utility scale and outcome weights &
$V=80$; $(w_{\mathrm{TP}},w_{\mathrm{FP}},w_{\mathrm{TN}},w_{\mathrm{FN}})
=(1.2,5.0,1.0,1.5)$ \\
Task-cost weights &
$\lambda_T=0.15$; $\lambda_c=\num{1e-4}$; $\eta_c=0.20$ \\
SN update cost &
$\lambda_s=4.0$ \\
Auxiliary losses &
$(w_g,w_r,w_m)=(1.0,0.5,0.2)$ for gate, route, and gate-margin losses;
$w_b=0.3$ for failure-boundary ranking \\
Exploration &
$\epsilon$-greedy; linearly decayed from $1.0$ to $0.05$ over
$3000$ steps \\
Training duration &
$80$ episodes for each load and arrival-process setting \\
Reported results &
Aggregated over three independent runs \\
\hline
\end{tabular}
\end{table}

The outcome weights prioritize accurate task selection. With $V=80$, the base
utilities before the secondary task-cost term are $+96$ for TP, $-400$ for FP,
$+80$ for TN, and $-120$ for FN. Successful service is therefore preferred to
justified rejection ($w_{\mathrm{TP}}V>w_{\mathrm{TN}}V$), which discourages an
overly conservative rejection policy. A false positive receives a larger
penalty than a false negative
($w_{\mathrm{FP}}V>w_{\mathrm{FN}}V$) because a task that is accepted but
misses its deadline consumes limited queue and compute resources without
meeting the service objective. The positive TN reward favors rejecting tasks
that cannot be served under overload, whereas the FN penalty discourages
excessive rejection.

The scale $V=80$ makes TP/FP/TN/FN outcomes the primary learning signal. The
weight $\eta_c=0.20$ keeps the task-cost term secondary when selecting among
feasible accepted routes, preventing delay or transfer cost from dominating the
accept/reject decision. The auxiliary loss weights prioritize gate prediction,
route prediction, and gate-margin separation in decreasing order. The smaller
ranking weight $w_b=0.3$ regularizes failure-boundary ordering without
overriding the primary decision losses.
\begin{table}[htbp]
\renewcommand{\arraystretch}{1.05}
\caption{Agent architecture and reinforcement-learning parameters.}
\label{tab:hparams}
\centering
\scriptsize
\setlength{\tabcolsep}{3pt}
\begin{tabular*}{\columnwidth}{@{}p{2.15cm}@{\extracolsep{\fill}}p{2.75cm}@{\extracolsep{\fill}}p{2.85cm}@{}}
\hline
\bfseries Parameter & \bfseries EN & \bfseries SN \\
\hline
Semantic encoder & -- & $4\!\rightarrow\!32\!\rightarrow\!16\!\rightarrow\!2$; pretrained and frozen during joint RL \\
Decision network & Local/remote MLPs $32$--$32$; fusion 64 & Projection 32; GRU hidden size 32 \\
Action space & \{Reject, local, offload\} & \{send, no-send\} $\times$ \{20, 25, 30, 35, 40, 45, 50, 100, 200\}\,ms \\
Encoder pretraining & -- & 20 ns-3 collection episodes; 50 epochs; batch 128; learning rate \num{5e-4} \\
RL learning rate & \num{3e-4} & \num{3e-4} \\
Discount $\gamma$ & Terminal target$^{a}$ & 0.99 (semi-Markov) \\
Batch size & 32 & 16 \\
Replay buffer capacity & 4096 (uniform) & 4096 (uniform) \\
Target-network synchronization & Inactive$^{a}$ & Every 200 updates \\
Optimizer update frequency & Every 4 transitions & Every 2 transitions \\
Sequence length & -- & 2 SN decisions \\
Optimization & Smooth $L_1$; gradient clip 1.0 & Smooth $L_1$; gradient clip 1.0 \\
\hline
\multicolumn{3}{@{}p{\linewidth}@{}}{\emph{Note:} The EN target-network object is retained for compatibility with the general Double DQN trainer. Because EN targets are task-terminal, the EN discount and target synchronization are inactive; the effective discount and target synchronization listed above apply only to the SN target.}\\
\end{tabular*}
\end{table}

\subsubsection{Baselines}
\label{subsec:baselines}

We compare \textsc{CoSMO} with six baselines under identical workloads and the
same evaluation protocol:
\begin{itemize}
\item \textbf{All local (EN)}: Accepts every task and executes it at the EN
without offloading or adaptive status updates.

\item \textbf{All offload (SN)}: Accepts every task and forwards it to the SN for
remote execution.

\item \textbf{Local until AP full}: Executes tasks locally while EN-side
capacity is available and otherwise forwards them to the SN.

\item \textbf{AoI threshold}: Uses an analytically derived continuous AoI
threshold to schedule raw-state updates without semantic awareness or
RL~\cite{aoi_survey}.

\item \textbf{AoV gap}: Sends a raw-state update when the normalized gap between
the current and cached SN states exceeds a fixed threshold.

\item \textbf{AoCI online}: Follows the online Age of Changed Information
(AoCI) threshold policy~\cite{peng2024aoci}, which adapts a delay-aware
update threshold from observed content changes.
\end{itemize}

We also evaluate a hierarchical-decision ablation, denoted by
\textsc{CoSMO} (w/o Gate). This variant replaces the accept/reject gate and the
local/offload routing head with a single flat dueling Double DQN head that directly
chooses among the same three EN actions. We evaluate the ablation at
representative light-load, capacity-boundary, and overload points
$\lambda\in\{120,220,320\}$~tasks/s.

\subsubsection{Datasets}
\label{subsec:workload}

The workload is generated synthetically from the task-arrival and attribute
distributions in Table~\ref{tab:workload}. We use two arrival processes at the
same mean rate $\lambda$: Deterministic arrivals with constant interarrival
times and Poisson arrivals with exponential interarrival times. Deterministic
arrivals provide evenly spaced tasks for controlled load variation. Poisson
arrivals introduce random bursts and temporary queues. Together, they test the
policy under smooth and variable traffic at the same mean load.
The seven arrival rates are chosen relative to the aggregate EN+SN compute
capacity. Since the EN provides $2\times\num{18000}$ cycles/ms and the SN
provides $4\times\num{27000}$ cycles/ms, the aggregate compute budget is
$\num{144000}$ cycles/ms. With mean task demand
$\bar{c}=(\num{1.0e5}+\num{1.2e6})/2=\num{6.5e5}$ cycles, the reference service
rate is
\begin{equation}
\mu^\star
=
\frac{1000\times\num{144000}}{\num{6.5e5}}
\approx 221 \ \text{tasks/s}.
\end{equation}
Thus, $\lambda\in\{120,160,200,220,240,280,320\}$ tasks/s spans
$\rho=\lambda/\mu^\star\in[0.54,1.45]$, covering underload, the
capacity-boundary region, and overload. Deadlines are generated from the
minimum feasible path delay plus limited slack, so overload requires active
rejection rather than blind acceptance.

\begin{table}[!t]
\renewcommand{\arraystretch}{1.3}
\caption{Dataset and workload parameters.}
\label{tab:workload}
\centering
\footnotesize
\begin{tabular}{p{2.3cm}p{5.2cm}}
\hline
\bfseries Parameter & \bfseries Value \\
\hline
Arrival process & deterministic (constant interval), Poisson
  (exponential interval) \\
Mean rate $\lambda$ & $\{120,160,200,220,240,280,320\}$ tasks/s
  ($\rho \in [0.54, 1.45]$) \\
Input size & $\mathcal{U}(\num{20000},\num{50000})$ B \\
Compute demand & $\mathcal{U}(\num{1.0e5},\num{1.2e6})$ cycles \\
Deadline & \texttt{e2e\_lower\_bound}:
  $1.35 \cdot \min(\text{local},\text{offload})$ path delay
  $+\,\mathcal{U}(10,30)$ ms, clipped to $[60,220]$ ms \\
\hline
\end{tabular}
\end{table}

\subsubsection{Evaluation Metrics}
\label{subsec:eval}

All learned-policy evaluations use greedy actions ($\epsilon=0$). Final metrics
for each run are computed from a five-episode ns-3 evaluation. Reported values
are means over three runs, and error bars show two-sided $95\%$ confidence
intervals. Arrivals before \SI{2000}{ms} are excluded as warm-up. Requests that
do not reach the EN within \SI{5}{s} after their deadlines receive zero reward
and are omitted from the TP/FP/TN/FN decision counts.
The primary metrics are:
\begin{itemize}
\item System success rate: The fraction of measured tasks completed
  before their deadlines.
\item Decision accuracy: $(\mathrm{TP}+\mathrm{TN})/N_{\mathrm{dec}}$,
  where $N_{\mathrm{dec}}$ is the number of tasks for which the EN makes a
  valid decision. Under overload, a rejection may be forced by limited
  compute capacity rather than by an incorrect task-selection decision. We
  therefore define a common capacity-based rejection quota:
  \begin{equation}
  \label{eq:quota}
  q = \left\lceil
  \frac{
      \max \left(0, D_{\mathrm{meas}}-\mu_{\mathrm{cyc}} W_{\mathrm{svc}}\right)
  }{\bar{c}}
  \right\rceil .
  \end{equation}
  The numerator is the demand that exceeds the compute service available
  during the service window; dividing by the mean per-task demand converts
  this excess into an estimated number of tasks. Specifically,
  $D_{\mathrm{meas}}$ is the measured compute demand in CPU cycles,
  $\bar{c}=D_{\mathrm{meas}}/N_{\mathrm{dec}}$ is the mean per-task demand,
  $\mu_{\mathrm{cyc}}=2\times\num{18000}+4\times\num{27000}
  =\num{144000}$ cycles/ms is the aggregate EN+SN service rate, and
  $W_{\mathrm{svc}}$ is the service window. A rejected task is normally a false
  negative when it could have been served on time. Under overload, however, the
  system may have to reject tasks because its capacity is insufficient. We
  therefore count up to $q$ such rejections as load-justified true negatives.
  The same quota construction is used for every method.
\item Update-cost rate: For each SN update decision, sending an update incurs
  cost $c_0 d_{\mathrm{upd}}$, while no-send incurs zero cost. We set $c_0=0.5$;
  \textsc{CoSMO} uses $d_{\mathrm{upd}}=2$ for semantic updates, and raw-state
  baselines use $d_{\mathrm{upd}}=4$. Here $c_0$ is the unit update-cost
  coefficient, $d_{\mathrm{upd}}$ denotes the update payload dimension, and
  the subscript $n$ indexes the $n$-th SN update decision. The normalized rate is
  \begin{equation}
  \label{eq:update_cost_rate}
  R_{\mathrm{upd}} = \frac{
        \sum_n c_0 d_{\mathrm{upd},n}
        \mathbbm{1}_{\{\text{send}_n=1\}}
      }{
        N_{\mathrm{task}}/\lambda
        }.
  \end{equation}
  In~\eqref{eq:update_cost_rate}, $\mathbbm{1}_{\{\text{send}_n=1\}}$ is an
  indicator that equals one if the $n$-th decision sends an update and zero
  otherwise, $N_{\mathrm{task}}$ is the number of generated tasks, and
  $\lambda$ is the arrival rate in tasks/s, so $N_{\mathrm{task}}/\lambda$
  approximates the workload duration in seconds.
\end{itemize}

\subsection{Results}
\label{sec:results}
\subsubsection{Capacity-Aware Decision Accuracy}
\label{subsec:res_decacc}

\begin{figure}[htbp]
\centering
\subfloat[Deterministic]{\includegraphics[width=0.49\columnwidth]{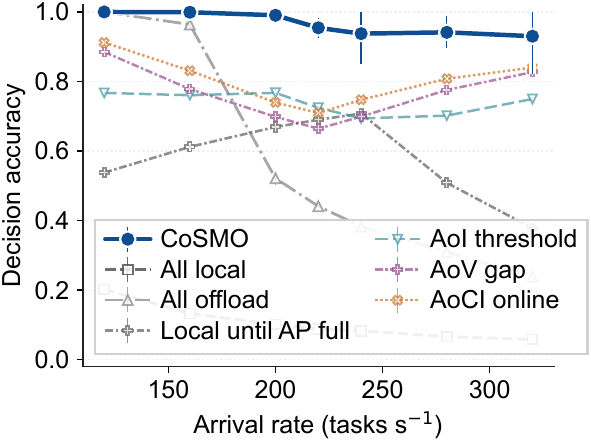}\label{fig:decacc_det}}\hfil
\subfloat[Poisson]{\includegraphics[width=0.49\columnwidth]{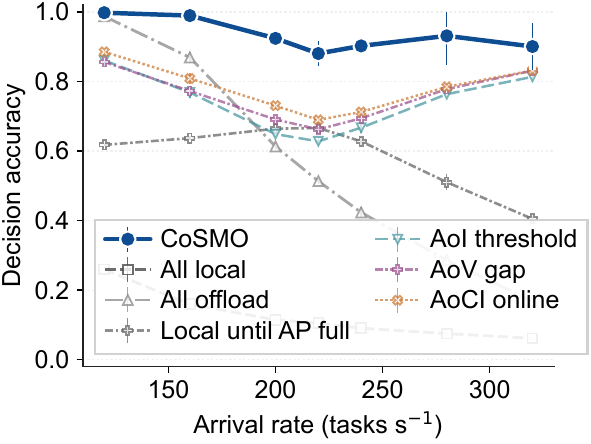}\label{fig:decacc_poi}}
\caption{Capacity-aware decision accuracy under deterministic and Poisson
arrivals.}
\label{fig:decacc}
\end{figure}

\figurename~\ref{fig:decacc} compares capacity-aware decision accuracy across
the complete load sweep. At $\lambda\leq\SI{200}{tasks/s}$
($\rho\leq0.90$), \textsc{CoSMO} achieves $0.990$--$1.000$ under
deterministic arrivals and $0.924$--$0.998$ under Poisson arrivals. At the
capacity boundary ($\lambda=\SI{220}{tasks/s}$), the corresponding values are
$0.954$ and $0.880$. Under strict overload
($\lambda\in\{240,280,320\}$~tasks/s), its accuracy remains
$0.930$--$0.941$ under deterministic arrivals and $0.901$--$0.931$ under
Poisson arrivals. For example, at \SI{320}{tasks/s} under Poisson arrivals,
decision accuracy remains $0.901$, although the system success rate is
$0.620$.

The rejection-capable baselines achieve lower decision accuracy under
overload. Across the three strict-overload points, \textsc{CoSMO}'s relative
gain over the best-performing competing method at each load ranges from
$10.7\%$ to $25.5\%$ under deterministic arrivals and from $8.3\%$ to
$26.7\%$ under Poisson arrivals. Averaged over these points, the corresponding
gains are $17.6\%$ and $17.9\%$. At the Poisson capacity boundary,
\textsc{CoSMO} achieves $0.880$, compared with $0.690$ for AoCI online, which
corresponds to a $27.7\%$ relative improvement. At \SI{320}{tasks/s}, the
three Poisson rejection-capable baselines range from $0.813$ to $0.832$.

Overall, \textsc{CoSMO} maintains high capacity-aware decision accuracy
from light load to deep overload and consistently outperforms the competing
baselines. These results show that its task-selection accuracy is maintained
across both arrival processes and the evaluated range of offered loads.

\subsubsection{On-Time Task Completion}
\label{subsec:res_success}

\begin{figure}[htbp]
\centering
\subfloat[Deterministic]{\includegraphics[width=0.49\columnwidth]{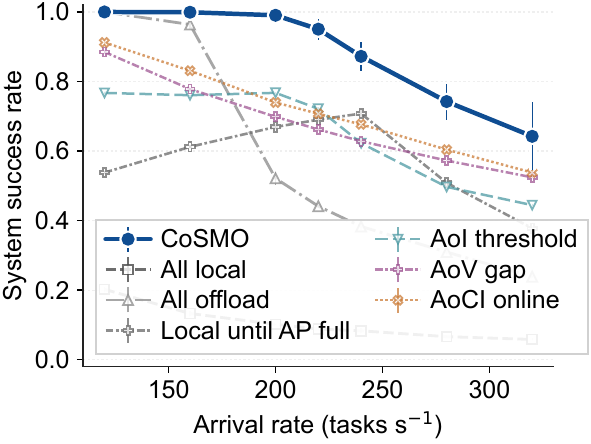}\label{fig:success_det}}\hfil
\subfloat[Poisson]{\includegraphics[width=0.49\columnwidth]{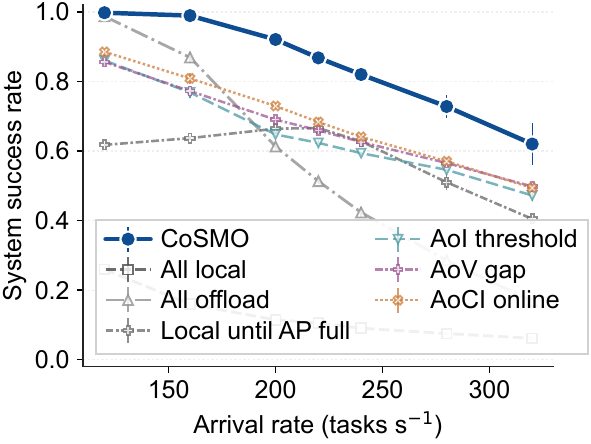}\label{fig:success_poi}}
\caption{System success rate under deterministic and Poisson arrivals.}
\label{fig:success}
\end{figure}

To compare on-time task completion across load conditions,
\figurename~\ref{fig:success} reports the system success rate, defined as the
fraction of measured tasks completed before their deadlines.

\textsc{CoSMO} attains the highest success rate at every evaluated load under
both arrival processes. Its deterministic success rate decreases from $1.000$
at \SI{120}{tasks/s} to $0.642$ at \SI{320}{tasks/s}; the corresponding
Poisson values are $0.998$ and $0.620$. The aggregate EN+SN compute capacity is
approximately $\mu^\star=221$ tasks/s. Therefore, at the strict-overload points
$\lambda\in\{240,280,320\}$ tasks/s, the capacity-limited upper bound is
$\mu^\star/\lambda$. \textsc{CoSMO} achieves $93\%$--$94\%$ of this bound
under deterministic arrivals and $89\%$--$92\%$ under Poisson arrivals. The
lower Poisson ratios are consistent with greater queueing variability from
bursty arrivals at the same mean load.

The fixed baselines degrade as the load increases because their static rules
either rely on a single execution site or cannot reject tasks under overload.
Under Poisson arrivals, All offload decreases from $0.987$ at
\SI{120}{tasks/s} to $0.168$ at \SI{320}{tasks/s} as the remote path
saturates. All local decreases from $0.259$ to $0.061$ because it can use only
EN compute. Local until AP full uses both execution sites but has no rejection
control and achieves only $0.405$ at the heaviest load. The rejection-capable
heuristics (AoV gap, AoI threshold, and AoCI online) outperform these fixed
policies under overload but remain below \textsc{CoSMO}.

Across all seven Poisson loads, \textsc{CoSMO}'s relative improvement over the
best-performing competing method at each load averages $21.2\%$. From the
capacity boundary through deep overload ($\lambda\geq\SI{220}{tasks/s}$), the
improvement ranges from $24.5\%$ to $28.0\%$. Its mean across the three
strict-overload points is $26.7\%$. The deterministic sweep yields a mean
relative improvement of $18.6\%$. At \SI{320}{tasks/s} under Poisson arrivals,
\textsc{CoSMO} achieves $0.620$, compared with $0.498$ for AoV gap, the
best-performing competing method. This difference corresponds to a $24.5\%$
improvement.

\textsc{CoSMO} matches or exceeds every competing method in on-time
completion rate at every load under both arrival processes and remains close
to the capacity-limited upper bound under overload. Together with the preceding
decision-accuracy results, these findings support \textsc{CoSMO}'s intended
framework-level behavior. Across the evaluated settings, the complete
closed-loop design coordinates SN semantic updates with EN task decisions while
maintaining accurate task selection and high on-time completion.

\subsubsection{Status-Update Efficiency}
\label{subsec:res_updcost}

\begin{figure}[htbp]
\centering
\subfloat[Deterministic]{\includegraphics[width=0.49\columnwidth]{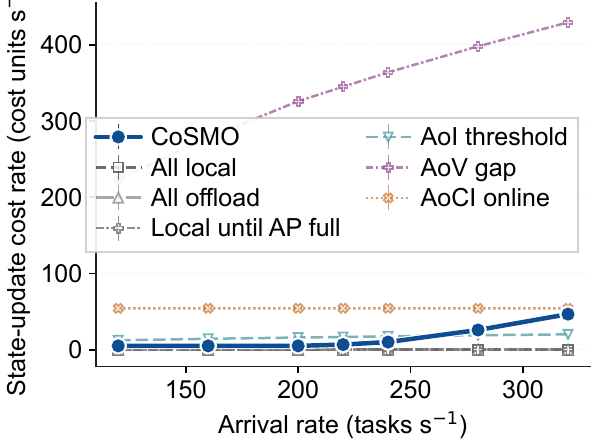}\label{fig:updcost_det}}\hfil
\subfloat[Poisson]{\includegraphics[width=0.49\columnwidth]{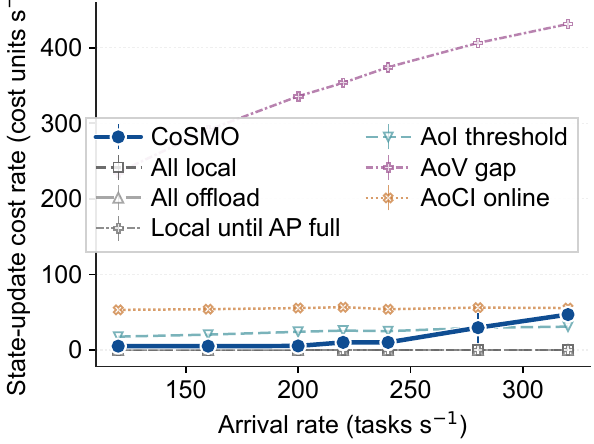}\label{fig:updcost_poi}}
\caption{Update-cost rate under deterministic and Poisson arrivals.}
\label{fig:updcost}
\end{figure}

To quantify status-update expenditure, we next compare the update-cost rate
defined in~\eqref{eq:update_cost_rate}. Methods that never transmit status
updates have zero update cost in \figurename~\ref{fig:updcost}.

Under deterministic arrivals, \textsc{CoSMO}'s update-cost rate increases from
$5.10$ cost units/s at \SI{120}{tasks/s} to $46.71$ cost units/s at
\SI{320}{tasks/s}. Under Poisson arrivals, it increases from $5.16$ to $46.96$
cost units/s over the same range. AoV gap incurs higher update cost, increasing
from $223.45$ to $429.53$ cost units/s under deterministic arrivals and from
$235.06$ to $431.40$ cost units/s under Poisson arrivals. AoCI online remains
within $53.18$--$56.68$ cost units/s and is also more expensive than
\textsc{CoSMO} throughout the sweep. At \SI{320}{tasks/s}, \textsc{CoSMO}
reduces update cost relative to AoV gap by $89.1\%$ under either arrival
process. Relative to AoCI online, the reductions are $14.1\%$ under
deterministic arrivals and $15.5\%$ under Poisson arrivals.

Among the policies that transmit status updates, AoI threshold has the lowest
mean update-cost rate at the two heaviest loads. From
$\lambda=\SI{120}{tasks/s}$ to $\lambda=\SI{320}{tasks/s}$, its rate increases
from $12.39$ to $20.19$ cost units/s under deterministic arrivals and from
$17.73$ to $30.97$ cost units/s under Poisson arrivals. At
$\lambda=\SI{320}{tasks/s}$, \textsc{CoSMO}'s corresponding rates are $46.71$
(deterministic) and $46.96$ (Poisson) cost units/s. These rates are $131.3\%$
and $51.6\%$ higher than those of AoI threshold. Across the separately trained
load settings, the learned SN policies allocate little update cost at light
load and more under overload. The SN controls this expenditure by deciding both
whether to send an update and when to make the next update decision, rather
than following a fixed update period.

Together, the update-cost and task metrics characterize \textsc{CoSMO}'s
closed-loop tradeoff. The load-specific learned policies keep update cost low
at light load and raise it under overload. At
$\lambda=\SI{320}{tasks/s}$, \textsc{CoSMO} incurs higher update cost than AoI
threshold but achieves higher system success rate and decision accuracy under
both arrival processes. It also outperforms the best-performing competitor for each
task metric. The additional update effort therefore accompanies better
decision quality and on-time completion under deep overload.

\subsubsection{Training-Reward Stabilization}
\label{subsec:res_reward}

\begin{figure}[htbp]
\centering
\subfloat[Deterministic]{\includegraphics[width=0.49\columnwidth]{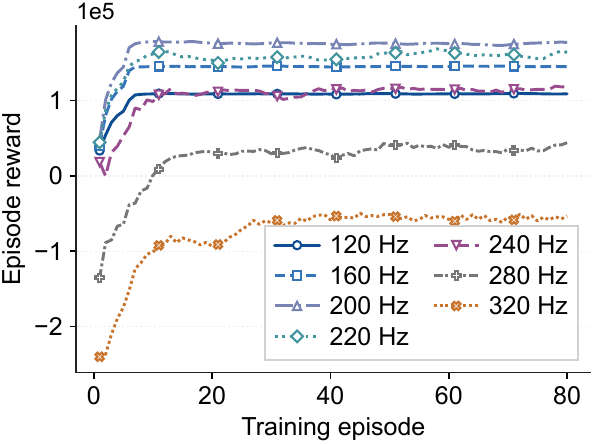}\label{fig:reward_det}}\hfil
\subfloat[Poisson]{\includegraphics[width=0.49\columnwidth]{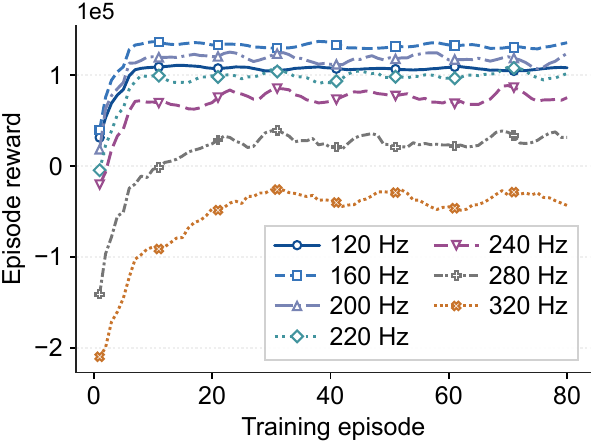}\label{fig:reward_poi}}
\caption{Training-reward trajectories under deterministic and Poisson
arrivals.}
\label{fig:reward}
\end{figure}

To assess how quickly the training reward stabilizes across load and
arrival-process settings, we examine its episode-wise trajectory.
\figurename~\ref{fig:reward} shows five-episode moving averages of the training
reward over three runs.

We use $t_{90}$ as a descriptive reward-stabilization indicator. For each run,
it is the first episode at which the smoothed reward reaches $90\%$ of the
change from its initial value to its final-ten-episode mean and remains beyond
that level for five episodes. The median $t_{90}$ is $6$--$7$ episodes for all
loads up to \SI{220}{tasks/s}. Under deterministic overload, the medians are
$11$, $13$, and $13$ episodes at $\lambda\in\{240,280,320\}$~tasks/s. Under
Poisson arrivals, they are $18$, $17$, and $20$ episodes. Thus, the median
$t_{90}$ does not exceed $20$ episodes in any evaluated setting.

\subsubsection{Hierarchical Decision Ablation}
\label{subsec:res_gate_ablation}

To isolate the hierarchical decision structure, we replace the accept/reject
gate and local/offload routing head with a single flat dueling Double DQN head that
selects the same three actions. The remaining architecture, training budget,
data, checkpoint selection, and evaluation protocol are unchanged. We compare this
variant, denoted as \textsc{CoSMO} (w/o Gate), with \textsc{CoSMO} using paired
runs at $\lambda\in\{120,220,320\}$~tasks/s.

Table~\ref{tab:gate_ablation} summarizes the paired success rate and
decision-accuracy results. At \SI{120}{tasks/s}, the two policies have nearly
identical mean outcomes.
At the capacity boundary, \textsc{CoSMO}'s mean success rate and decision accuracy
are higher than those of \textsc{CoSMO} (w/o Gate) by $1.0\%$ and $1.3\%$
under deterministic arrivals and by $1.5\%$ and $1.0\%$ under Poisson
arrivals. The relative gains are larger in deep overload. At
\SI{320}{tasks/s}, they are $13.4\%$ in success rate and $9.9\%$ in decision
accuracy under deterministic arrivals, and $8.8\%$ and $2.6\%$ under Poisson
arrivals. As detailed in Table~\ref{tab:gate_ablation_detailed}, at the same
Poisson load, the hierarchical policy achieves $4.7\%$ higher mean
accepted-task success rate and $4.0\%$ higher mean route accuracy. It also reduces
the mean false-positive rate by $18.9\%$, from $0.123$ to $0.099$, while using
comparable mean update cost.

\begin{table}[!t]
\renewcommand{\arraystretch}{1.08}
\setlength{\tabcolsep}{2.6pt}
\caption{Hierarchical gate--route ablation (mean $\pm$ standard deviation).}
\label{tab:gate_ablation}
\centering
\scriptsize
\begin{tabular}{c l c c}
\hline
$\boldsymbol{\lambda}$ & \bfseries Method &
\shortstack{\bfseries Success\\\bfseries rate} &
\shortstack{\bfseries Decision\\\bfseries accuracy} \\
\hline
\multicolumn{4}{c}{\itshape Deterministic arrivals} \\
\hline
120 & \textsc{CoSMO} & $1.000{\pm}0.000$ & $1.000{\pm}0.000$ \\
    & \textsc{CoSMO} (w/o Gate) & $1.000{\pm}0.000$ & $1.000{\pm}0.000$ \\
220 & \textsc{CoSMO} & $0.950{\pm}0.012$ & $0.954{\pm}0.012$ \\
    & \textsc{CoSMO} (w/o Gate) & $0.941{\pm}0.006$ & $0.943{\pm}0.008$ \\
320 & \textsc{CoSMO} & $0.642{\pm}0.039$ & $0.930{\pm}0.043$ \\
    & \textsc{CoSMO} (w/o Gate) & $0.566{\pm}0.021$ & $0.847{\pm}0.035$ \\
\hline
\multicolumn{4}{c}{\itshape Poisson arrivals} \\
\hline
120 & \textsc{CoSMO} & $0.998{\pm}0.001$ & $0.998{\pm}0.001$ \\
    & \textsc{CoSMO} (w/o Gate) & $0.999{\pm}{<}0.001$ & $0.999{\pm}{<}0.001$ \\
220 & \textsc{CoSMO} & $0.868{\pm}0.010$ & $0.880{\pm}0.014$ \\
    & \textsc{CoSMO} (w/o Gate) & $0.856{\pm}0.004$ & $0.872{\pm}0.006$ \\
320 & \textsc{CoSMO} & $0.620{\pm}0.024$ & $0.901{\pm}0.027$ \\
    & \textsc{CoSMO} (w/o Gate) & $0.570{\pm}0.044$ & $0.877{\pm}0.049$ \\
\hline
\end{tabular}
\end{table}

\begin{table}[!t]
\renewcommand{\arraystretch}{1.10}
\setlength{\tabcolsep}{2.5pt}
\caption{Hierarchical gate--route ablation metrics under Poisson arrivals
(mean $\pm$ standard deviation).}
\label{tab:gate_ablation_detailed}
\centering
\scriptsize

\begin{tabular}{@{}c l c c c@{}}
\hline
$\boldsymbol{\lambda}$ & \bfseries Method &
\bfseries FP $\downarrow$ &
\bfseries FN $\downarrow$ &
\bfseries Reject \\
\hline
\multicolumn{5}{c}{\itshape Gate-decision metrics} \\
\hline
\multirow{2}{*}{120}
& \textsc{CoSMO}
& $0.002{\pm}0.001$
& $0.000{\pm}0.000$
& $0.000{\pm}0.000$ \\
& \textsc{CoSMO} (w/o Gate)
& $0.001{\pm}{<}0.001$
& $0.000{\pm}0.000$
& $0.000{\pm}0.000$ \\
\hline
\multirow{2}{*}{220}
& \textsc{CoSMO}
& $0.109{\pm}0.025$
& $0.011{\pm}0.011$
& $0.023{\pm}0.016$ \\
& \textsc{CoSMO} (w/o Gate)
& $0.111{\pm}0.018$
& $0.017{\pm}0.011$
& $0.033{\pm}0.016$ \\
\hline
\multirow{2}{*}{320}
& \textsc{CoSMO}
& $0.099{\pm}0.027$
& $0.000{\pm}0.000$
& $0.280{\pm}0.004$ \\
& \textsc{CoSMO} (w/o Gate)
& $0.123{\pm}0.049$
& $0.000{\pm}0.000$
& $0.307{\pm}0.016$ \\
\hline
\end{tabular}

\vspace{1.2mm}

\begin{tabular}{@{}c l c c c@{}}
\hline
$\boldsymbol{\lambda}$ & \bfseries Method &
\shortstack{\bfseries Accepted-task\\
            \bfseries success rate $\uparrow$} &
\shortstack{\bfseries Route\\
            \bfseries accuracy $\uparrow$} &
\shortstack{\bfseries Episode\\
            \bfseries update cost $\downarrow$} \\
\hline
\multicolumn{5}{c}{\itshape Task-routing and update metrics} \\
\hline
\multirow{2}{*}{120}
& \textsc{CoSMO}
& $0.998{\pm}0.001$
& $0.998{\pm}0.001$
& $51{\pm}0$ \\
& \textsc{CoSMO} (w/o Gate)
& $0.999{\pm}{<}0.001$
& $0.999{\pm}{<}0.001$
& $51{\pm}0$ \\
\hline
\multirow{2}{*}{220}
& \textsc{CoSMO}
& $0.889{\pm}0.024$
& $0.911{\pm}0.013$
& $101{\pm}27$ \\
& \textsc{CoSMO} (w/o Gate)
& $0.885{\pm}0.017$
& $0.917{\pm}0.007$
& $105{\pm}18$ \\
\hline
\multirow{2}{*}{320}
& \textsc{CoSMO}
& $0.862{\pm}0.036$
& $0.895{\pm}0.025$
& $471{\pm}36$ \\
& \textsc{CoSMO} (w/o Gate)
& $0.823{\pm}0.068$
& $0.861{\pm}0.059$
& $468{\pm}69$ \\
\hline
\end{tabular}
\end{table}

\begin{figure}[!t]
\centering
\subfloat[Deterministic arrivals]{\includegraphics[width=0.49\linewidth]{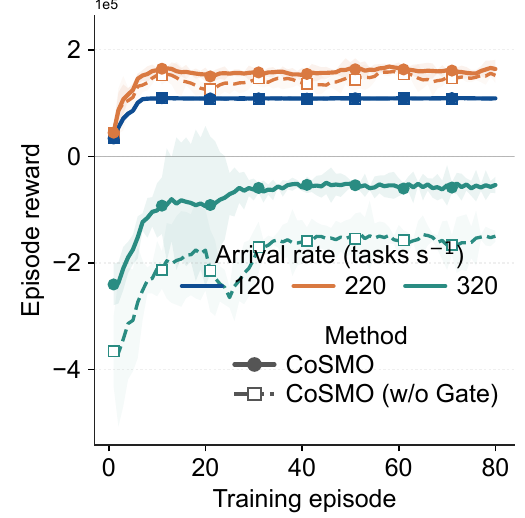}\label{fig:gate_ablation_reward_det}}\hfil
\subfloat[Poisson arrivals]{\includegraphics[width=0.49\linewidth]{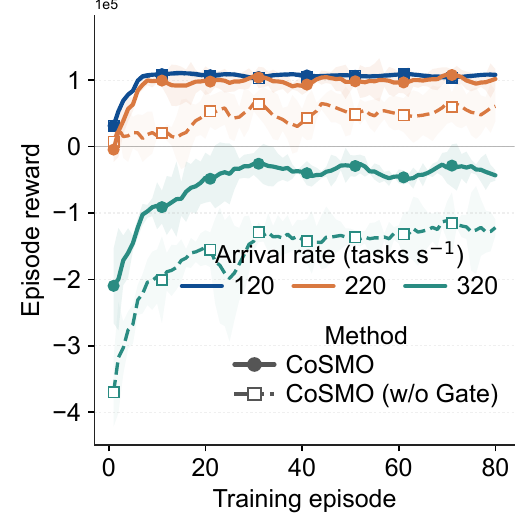}\label{fig:gate_ablation_reward_poi}}
\caption{Training-reward trajectories with and without the hierarchical
gate--route factorization.}
\label{fig:gate_ablation_reward}
\end{figure}

\figurename~\ref{fig:gate_ablation_reward} shows that the two policies have
similar training-reward trajectories at light load. Under deep overload, the
hierarchical policy stabilizes earlier and reaches a higher reward level for
both arrival processes. Together with the task-outcome results, the mean trends
suggest that the gate--route factorization is the most useful when accept/reject
and routing decisions become tightly coupled under overload. In particular,
under Poisson arrivals, where the instantaneous workload fluctuates more
drastically around the same mean arrival rate, the hierarchical policy retains
its convergence advantage over the flat action head, indicating more stable
training behavior under workload variability.

\section{Conclusion}
\label{sec:conclusion}

We proposed \textsc{CoSMO}, a cooperative semantic-aware status-updating and
selective-offloading framework for decisions based on stale remote state.
Across separately trained deterministic and Poisson load settings,
\textsc{CoSMO} maintained high capacity-aware decision accuracy under overload
and achieved the highest on-time completion rate. Its learned policies did not
minimize update cost at every load. Instead, they used low update cost at light
load and more under overload. This tradeoff is accompanied by higher task
metrics than AoI threshold and lower update cost than AoV gap. The descriptive ablation
trends associate the hierarchical gate--route structure with higher task
metrics and earlier reward stabilization under deep overload. End-to-end comparisons support the complete decision-oriented co-design within
the evaluated topology. The gate ablation isolates only the hierarchical EN
factorization; the remaining gains are reported at the
integrated-framework level rather than attributed to any individual module.

The evaluation is limited to a controlled single-EN/single-SN topology.
Fixed-rate backhaul and a fixed modulation and coding scheme isolate the
coupling of status staleness with accept/reject and execution-routing decisions
but do not test transfer across topologies. Future work will consider multiple
ENs and SNs, time-varying backhaul, interference-coupled access, and policy
transfer across topologies.

\bibliographystyle{IEEEtran}
\bibliography{refs}

\end{document}